\documentclass[preprint]{aastex}

\usepackage{morefloats}
\usepackage{graphicx}

\shorttitle{}
\shortauthors{}

\begin{document}

\title{The Structure of the Local Interstellar Medium. VI. New \ion{Mg}{2}, \ion{Fe}{2}, and \ion{Mn}{2}  Observations Toward Stars Within 100 Parsecs}

\author{Craig Malamut\altaffilmark{1}, Seth Redfield\altaffilmark{1}, Jeffrey L. Linsky\altaffilmark{2}, Brian E. Wood\altaffilmark{3}, and Thomas R. Ayres\altaffilmark{4}}
\altaffiltext{1}{Department of Astronomy and Van Vleck Observatory,
Wesleyan University, Middletown, CT, 06459; {\tt
cmalamut@wesleyan.edu; sredfield@wesleyan.edu}}
\altaffiltext{2}{JILA, University of Colorado and NIST, Boulder, CO 80309}
\altaffiltext{3}{Naval Research Laboratory, Space Science Division, Washington, DC 20375, USA}
\altaffiltext{4}{CASA, University of Colorado, 389UCB Boulder, CO 80309-0389, USA}

\begin{abstract}

We analyze high-resolution spectra obtained with the Space Telescope Imaging Spectrograph onboard the \emph{Hubble Space Telescope} toward 34 nearby stars ($\leq$100 pc) to record \ion{Mg}{2}, \ion{Fe}{2} and \ion{Mn}{2} absorption due to the local interstellar medium (LISM). Observations span the entire sky, probing previously unobserved regions of the LISM. The heavy ions studied in this survey produce narrow absorption features that facilitate the identification of multiple interstellar components. We detected one to six individual absorption components along any given sight line, and the number of absorbers roughly correlates with the pathlength.  This high-resolution near-ultraviolet (NUV) spectroscopic survey was specifically designed for sight lines with existing far-UV (FUV) observations.  The FUV spectra include many intrinsically broad absorption lines (i.e., of low atomic mass ions) and often observed at medium resolution.  The LISM NUV narrow-line absorption component structure presented here can be used to more accurately interpret the archival FUV observations.  As an example of this synergy, we present a new analysis of the temperature and turbulence along the line of sight toward $\epsilon$ Ind.  The new observations of LISM velocity structure are also critical in the interpretation of astrospheric absorption derived from fitting the saturated \ion{H}{1} Lyman-$\alpha$ profile.  As an example, we reanalyze the spectrum of $\lambda$ And and find that this star likely does have an astrosphere.  Two stars in the sample that have circumstellar disks (49 Cet and HD141569) show evidence for absorption due to disk gas.  Finally, the substantially increased number of sight lines is used to test and refine the three-dimensional kinematic model of the LISM, and search for previously unidentified clouds within the Local Bubble.  We find that every prediction made by the \cite{redfield07lism4} kinematic model of the LISM is confirmed by an observed component in the new lines of sight.

\end{abstract}



\section{Introduction}
\label{introduction.s}


The interstellar medium (ISM) connects many fundamental areas of
astrophysics.  The morphology, density, and temperature of the ISM
control star formation \citep{evans99}; the dynamics of the ISM provides information on stellar winds \citep{linsky96,frisch95,muller06}; nearby electron density enhancements cause the scintillation of
distant radio sources \citep{linsky08}; the
ionization of the ISM yields clues to the interstellar
radiation field \citep{vallerga98}; and the chemical abundances and
enrichment of the ISM are tracers for the death of low-mass stars
as well as the supernovae of massive stars \citep{mccray79}.  Theoretical studies of the
phases of the ISM have produced classic works \citep[e.g.,][]{field69,mckee77},
presenting ideas which are still being analyzed and discussed today
\citep{heiles01}.

The local interstellar medium (LISM) is clearly important in the
context of the general ISM, as it provides the best opportunity to study in
great detail many of the physical processes that dictate its structure and evolution, such as gas dynamics,
pressure balance, and the effects of the radiation field.  Not only are these phenomena widespread in the ISM throughout our own Galaxy, but also in other galactic
environments even at high redshift \citep{frisch95,mckee98}.  LISM studies can also have a surprising impact on fields as disparate as astrobiology and geophysics.  For example, the interaction of the LISM with solar and stellar
winds controls the size and properties of the heliosphere and
astrospheres \citep{shapley21, begelman76, zank99,redfield06, wyman13}, which in turn affect cosmic ray fluxes in the associated planetary systems.  Finally, a densely sampled model of the LISM
could be used to predict and remove foreground contamination due to local
interstellar absorption in order to aid interpretation of spectra of
more distant targets in our Galaxy.

The Sun and nearby stars reside in a region of ionized material known as the Local Bubble (or Local Cavity). The first evidence for this arguably hot cavity came from color excess maps indicating a large pocket in the dust at its edge and observations of diffuse soft X-ray background observed across the entire sky (\citealt{frisch11} and references therein). The edge of the Local Bubble can be traced by the onset of \ion{Na}{1} and \ion{Ca}{2} absorption, indicators of colder material. This edge begins anywhere from 65 to 250 pc depending on the observed direction \citep{sfeir99,welsh10}. The initial carving of the Local Bubble was likely the result of stellar winds or supernova explosions.


Within the Local Bubble, isolated clouds of warm, partially ionized gas are observed \citep{redfield07lism4}, each distinguished by its own unique properties (e.g., density, temperature, projected velocity). The predominant strategy to study the LISM is to observe its absorption signatures against bright, nearby background sources. The shape and Doppler shift of absorption features offer insight into the nature of the ISM along each line of sight. 

Since resonance lines of common ions in the ISM  are formed mainly in the ultraviolet (UV), the advent of space-based high-resolution UV spectrographs, largely thanks to the {\it Hubble Space Telescope} (\emph{HST}), has made it possible to study the warm material in the LISM in unprecedented detail. The proximity of the LISM material permits detailed scrutiny currently impractical for longer distance scales; for sight lines of hundreds to thousands of parsecs, ISM absorptions are often blended and/or saturated. Conversely, by observing nearby stars, multiple LISM component absorption profiles are frequently fully resolved, allowing the identification and characterization of the constituent clouds.

The observation of heavier elements in warm clouds has proven to be a boon to the understanding of the structure of the LISM. Their relatively large masses reduce thermal broadening, and thus blending, allowing for more precise measurements of cloud velocities and straightforward identification of multiple structures along a line of sight. Of particular importance are \ion{Mg}{2} and \ion{Fe}{2}, which have high cosmic abundance and are the dominant ionization stages in the LISM \citep{slavin08}. Both produce multiple spectral lines that provide redundant measurements of each ion along a given line of sight. These heavy ions have been exploited extensively to characterize the global structure of the LISM.  \cite{genova90} used the {\it International Ultraviolet Explorer} (\emph{IUE}) to observe the \ion{Mg}{2} $h$ and $k$ lines of cool stars within 30 pc of the Sun.  ISM absorption superimposed on the \ion{Mg}{2} chromospheric emission profiles hinted at heterogeneities in the column density distribution, as well as unresolved clouds beyond the ``Local Cloud." Later studies using Goddard High Resolution Spectrograph (GHRS) onboard {\it HST}, identified the two nearest clouds --- the Local Interstellar Cloud (LIC) and the Galactic (G) Cloud --- and established a velocity vector with only $\sim$10 lines of sight \citep{lallement92,lallement95}.

\cite{redfield02} analyzed and compiled the (then) complete LISM sample of \ion{Mg}{2} and \ion{Fe}{2} observations taken at high resolution with {\it HST}'s GHRS and the Space Telescope Imaging Spectrograph (STIS).  \citet{redfield07lism4} used these data to develop a kinematic model of the LISM and to identify 15 distinct clouds each with a unique velocity vector. Observations of multiple ions and ionization levels in these clouds have enabled determinations of ionization structure \citep{wood02}; abundances and element depletions \citep{redfield04a}; and temperature and turbulence \citep{redfield04b}. Furthermore, increased numbers of sight lines have made it possible to examine the small-scale structure of the LIC \citep{redfield01}. The goal of the present study is to build on this historical body of data by adding a large number of observations of heavy ions along more distant sight lines, thereby extending and refining measurements of the LISM.


\begin{deluxetable}{llcccccccccl}
\rotate
\tablewidth{0pt}
\tabletypesize{\tiny}
\tablecaption{Parameters for Stars in the LISM SNAP Program\tablenotemark{a} \label{tab1}}
\tablehead{& & Spectral & $m_V$ & $v_R$ & $l$ & $b$ & Distance & S/N\tablenotemark{b} & S/N\tablenotemark{b} & S/N\tablenotemark{b} & Other \\
HD No. & Other Name & Type & (mag) & (km s$^{-1}$) & (deg) & (deg) & (pc) & (\ion{Mg}{2}) & (\ion{Fe}{2}) & (\ion{Mn}{2}) & Spectra}
\startdata
209100 & $\epsilon$ Ind & K5V & 4.833 & --40.4 & 336.2 & --48.0 & 3.62 & 29 & 6  & 5  & GHRS/Ech-A (Ly$\alpha$)\\
115617 & 61 Vir         & G5V & 4.74  & --8.5  & 311.9 & 44.1   & 8.56 & 18 & 6  & 7  & STIS/E140M E230M\\
114710 & $\beta$ Com    & G0V & 4.311 & 6.1    & 43.5  & 85.4   & 9.13 & 21 & 8  & 8  & FUSE\\
       & WD1620--391    & DA  & 10.974 & 43.2\tablenotemark{c}&341.5&7.3&13.2&6&7& 7  & GHRS/G160M, FUSE\\
72905  & $\pi^1$ UMa    & G1.5V & 5.706 & --12.0 & 150.6 & 35.7 & 14.4 & 22 & 8  & 7  & FUSE\\
217014 & 51 Peg         & G5V & 5.524 & --31.2 & 90.1  & --34.7 & 15.6 & 8  & 5  & 6  & STIS/G140M (Ly$\alpha$), FUSE\\
120136 & $\tau$ Boo     & F7V & 4.541 & --15.6 & 358.9 & 73.9   & 15.6 & 17 & 8  & 10 & STIS/G140M (Ly$\alpha$)\\
142373 & $\chi$ Her     & F9V & 4.672 & --55.4 & 67.7  & 50.3   & 15.9 & 8  & 8  & 16 & STIS/E140M \\
220140 & V368 Cep       & G9V & 7.622 & --16.8 & 118.5 & 16.9   & 19.2 & 19 & 5  & 2  & GHRS/G140M G160M G270M\\
97334  & MN UMa         & G0V & 6.476 & --2.6  & 184.3 & 67.3   & 21.9 & 16 & 5  & 3  & STIS/E140M E230M\\
       & WD1337+705     & DA  & 12.8  & 26     & 117.2 & 46.3   & 26.1 & 2  & 4  & 4  & STIS/G430M, FUSE\\
222107 & $\lambda$ And  & G8III--IV&3.975& 6.8 & 109.9 & --14.5 & 26.4 & 54 & 11 & 6  & GHRS/Ech-A (Ly$\alpha$), FUSE\\
180711 & $\delta$ Dra   & G9III & 3.188 & 24.8 & 98.7  & 23.0   & 29.9 & 20 & 5  & 5  & FUSE\\
12230  & 47 Cas         & F0V & 5.26  & --26   & 127.1 & 15.0   & 33.2 & 12 & 17 & 19 & GHRS/G140M, FUSE \\
163588 & $\xi$ Dra      & K2III & 3.867 & --26.4 & 85.2 & 30.2  & 34.5 & 23 & 4  & 3  & FUSE\\
216228 & $\iota$ Cep    & K0III & 3.621 & --12.6 & 111.1 & 6.2  & 35.3 & 24 & 5  & 5  & FUSE\\
93497  & $\mu$ Vel      & G5III & 2.818 & 6.2  & 283.0 & 8.6    & 35.9 & 39 & 11 & 8  & STIS/E140M, FUSE\\
149499 & V841 Ara       & K0V & 8.737 & --24.8 & 329.9 & --7.0  & 36.4 & 10 & 2  & 0  & STIS/E140M, FUSE\\
131873 & $\beta$ UMi    & K4III & 2.238 & 17.0 & 112.6 & 40.5   & 40.1 & 9  & 2  & 3  & FUSE\\
210334 & AR Lac         & G2IV & 6.203 & --34.6 & 95.6 & --8.3  & 42.8 & 13 & 5  & 5  & GHRS/G160M G270M, FUSE\\
28911  & HIP21267       & F5V & 6.619 & 35     & 183.4 & --22.6 & 44.7 & 12 & 6  & 10 & FUSE\\
28677  & 85 Tau         & F4V & 6.02  & 36     & 180.9 & --21.4 & 45.2 & 11 & 13 & 16 & FUSE\\
204188 & IK Peg         & A8  & 6.06  & --11.4 & 70.4  & --22.0 & 46.4 & 7  & 12 & 15 & GHRS/G160M, FUSE\\
       & WD0549+158     & DA  & 13.06 & 12.0   & 192.0 & --5.3  & 49\tablenotemark{d} & 4  & 5  & 5  & STIS/G140M G230M, FUSE\\
       & WD2004--605     & DA  & 13.14 & --26.5 & 336.6 & --32.9 & 58\tablenotemark{d}   & 3  & 4  & 5  & FUSE\\
9672   & 49 Cet         & A1V & 5.62  & 12.1\tablenotemark{e}&166.3&--74.8&59.4&24&37&37& FUSE\\
43940  & HR2265         & A2V  & 5.88 & 24.0\tablenotemark{f}&244.6&--22.4&61.9&19&29&30& FUSE\\*
137333 & $\rho$ Oct     & A2V & 5.57  & --11   & 307.0 & --23.0 & 66.1 & 14 & 25 & 31 & FUSE\\*
       & WD1631+781     & DA  & 13.03 & \nodata &111.3& 33.6   & 67\tablenotemark{d}   & 0  & 0  & 0 & FUSE\\*
3712   & $\alpha$ Cas   & K0II--III &2.377&--4.3& 121.4 & --6.3 & 70.0 & 21 & 6  & 5  & FUSE\\*
149382 & HIP81145       & B5  & 8.872 & 3      & 11.8  & 27.9   & 73.9 & 21 & 24 & 20 & FUSE\tablenotemark{g}\\*
       & WD0621--376     & DA  & 11.99 & 40.5\tablenotemark{c}&245.4&--21.4&78\tablenotemark{d}&8&9& 9  & FUSE \\*
75747  & RS Cha         & A7V & 6.02  & 26.0   & 292.6 & --21.6 & 92.9 & 9  & 17 & 18 & STIS/E230M, FUSE\\*
       & IX Vel         & O9  & 9.503 & 20     & 264.9 & --7.9  & 96.7 & 9  & 10 & 10 & STIS/E140M, FUSE\\*
141569 & HIP77542       & B9  & 7.143 & --7.6\tablenotemark{h}&4.2&36.9&116&11&14&13 & FUSE\\*
149730 & R Ara          & B9IV/V & 6.73 & $\pm$100\tablenotemark{i}&330.4&--6.8&124&14&20&21&FUSE
\enddata
\tablenotetext{a}{All stellar parameters taken from the SIMBAD database unless otherwise stated.}
\tablenotetext{b}{S/N calculated over 10 km~s$^{-1}$ bins on either side of any LISM absorption.}
\tablenotetext{c}{\cite{holberg98}}
\tablenotetext{d}{\cite{vennes97}}
\tablenotetext{e}{\cite{hughes08}}
\tablenotetext{f}{\cite{gontcharov06}}
\tablenotetext{g}{Planned FUSE observations that were not completed before the end of the mission.}
\tablenotetext{h}{\cite{dent05}}
\tablenotetext{i}{\cite{reed10}}
\end{deluxetable}

\section{Observations}
\label{observations.s}

\subsection{A SNAP Survey with STIS}
\label{a.snapshot.survey.with.stis.ss}

STIS is a powerful multi-mode UV/optical spectrograph, whose NUV (1600--3100 \AA) high-resolution echelle capability is well suited for detecting the narrow absorption lines of heavy ions in the warm, partially ionized LISM. For the present ``SNAPshot'' program -- designed to provide short observations to fill gaps in the {\it HST} schedule -- the E230H/2713~\AA\ setting was used exclusively.  The spectral resolving power is $R \equiv \frac{\lambda}{\Delta\lambda} \sim$ 114,000 with a simultaneous range of $\sim$200 \AA.  This setting captures several lines of interest, namely, \ion{Mg}{2} (2796.3543 {\AA} and 2803.5315 {\AA}), \ion{Fe}{2} (2586.6500 {\AA} and 2600.1729 {\AA}), and \ion{Mn}{2} (2594.499 {\AA} and 2606.462 {\AA})\footnote{Vacuum rest wavelengths from \citet{morton03}}. 

\begin{figure}[!ht]
\centering
\centerline{\includegraphics{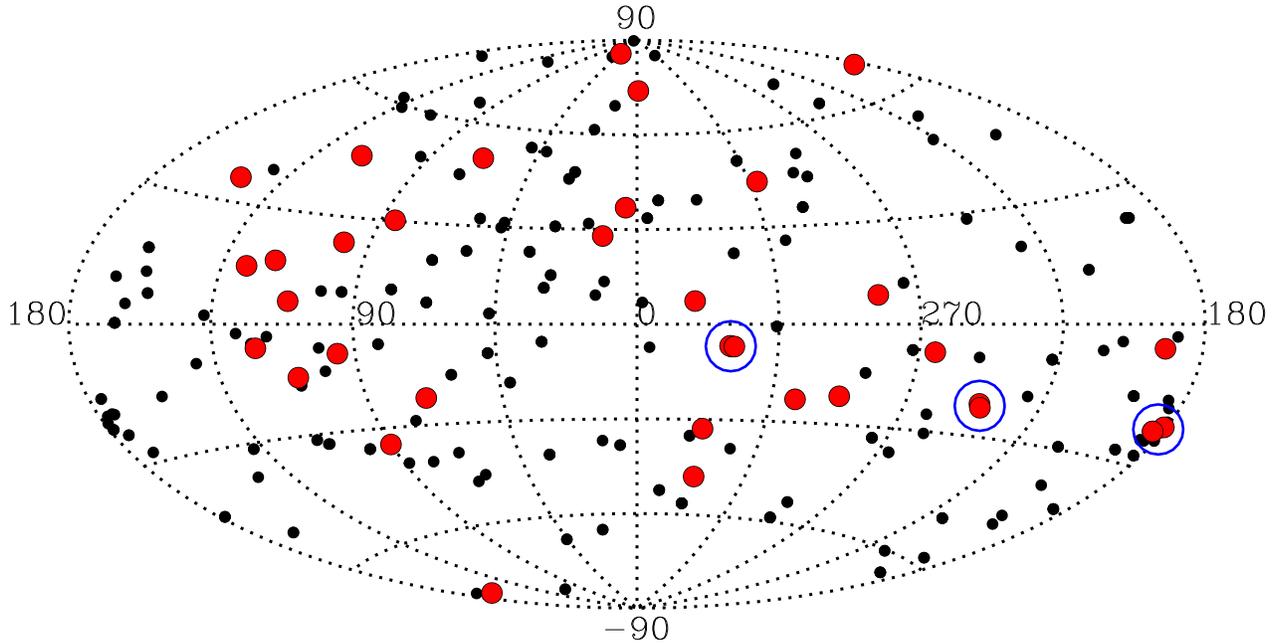}}
\caption{Map in Galactic coordinates of all sight lines for which LISM spectra have been obtained. Large, red circles indicate sight lines added by the SNAP survey and described in this work.  Filled black circles indicate sight lines previously analyzed by \citet{redfield07lism4}.  The three pairs of stars (circled) with small angular separation offer an opportunity to probe small-scale structure in the LISM clouds (see Section~\ref{secsss}). \label{galloc}}
\end{figure}

The LISM SNAP survey recorded spectra for 34 sight lines toward stars within $\sim$100 pc (Figure \ref{galloc}). While the original sample was restricted to targets $\leq$100 pc, the recent reexamination of {\it Hipparcos} parallaxes by \citet{vanleeuwen07} resulted in a revision of distance beyond 100 pc for two of our targets (HD141569 and R Ara).  High-resolution NUV observations of nearby stars are ideally suited for an {\it HST} SNAP program because suitable targets are broadly distributed across the sky and the limited partial orbit exposure time can still achieve a S/N $ > 10$.  In contrast, similar observations in the FUV require significantly more time, and must often be obtained at lower spectral resolution.  This SNAP survey sought to optimize the LISM sample by taking high-resolution NUV observations of targets that {\it already} had the more time-consuming FUV spectra in the archive (from {\it HST}'s GHRS or STIS, or the {\it Far-Ultraviolet Spectroscopic Explorer} [{\it FUSE}]).  While the original FUV observations might not have been taken with LISM analysis in mind, they can be utilized in combination with the new NUV spectra to make a comprehensive analysis of the LISM absorption along the particular line of sight.  The SNAP NUV spectra have average S/N of 15, 10, and 10 for the \ion{Mg}{2}, \ion{Fe}{2}, and \ion{Mn}{2} lines, respectively, over a 10 km~s$^{-1}$ range adjacent to any LISM absorption feature.  Table~\ref{tab1} provides a full list of the observed targets.

Two additional sight lines (also listed in Table~\ref{tab1}) were observed but did not result in spectra with detectable LISM absorption.  One of the targets, $\beta$ UMi, shows a dramatic P Cygni wind profile which prevents a confident placement of the continuum and obscures ISM features.  The other target, WD1631+781, was not analyzed further owing to anomalously low S/N.



\subsection{Data Reduction}

Initially, we processed the SNAP NUV spectra through the most recent available version of the {\tt calstis} pipeline.  We then utilized a second pipeline to perform a series of finer calibrations, based on protocols developed for the ``StarCAT'' catalog of STIS echelle spectra of 545 stellar objects \citep{ayres10}.  The StarCAT pipeline corrects each exposure for wavelength-dependent distortions in the STIS dispersion relations, specialized for the particular echelle setting, concatenates the $\sim$40 echelle orders captured in the exposure into a coherent one-dimensional spectrum, coadds spectra of similar type (i.e., subexposures in a given setting), and splices together neighboring spectra into a uniform tracing covering the available spectral range.

\section{Spectral Analysis}
\label{spectral.analysis.s}

We measured absorption features using the same procedures described in \cite{redfield02}. The stellar emission, or continuum, across the ISM absorption was modeled by a least-squares polynomial of order 1 to 10 applied to the surrounding region. Flat or simple continua  (e.g., WD1620-391 in Figure \ref{vp2}) could be fitted with lower order polynomials, while more complex examples (e.g., $\iota$ Cep in  Figure \ref{vp8}) required higher order polynomials. A Marquart $\chi^{2}$ minimization approach was applied to the maximum number of Voigt components that were statistically justified by an F-test. Each discrete Voigt profile represents an individual interstellar component. The uncertainty for each parameter was based on a Monte Carlo analysis.

For a particular ion, the fitting procedure was performed twice: once treating each observed line of the multiplet individually; and a second time for a simultaneous fit. The simultaneous fitting was restricted to transitions within the same ion.  Fitting \ion{Fe}{2} and \ion{Mg}{2} simultaneously, for example, would require that the ions are well mixed and in thermodynamic equilibrium.  While these likely are good assumptions, treating each ion individually not only allows a test of these assumptions, but also enables other independent comparisons of the two ions along each line of sight.
	
\begin{figure}
\centering
\figurenum{2a}
\centerline{\includegraphics{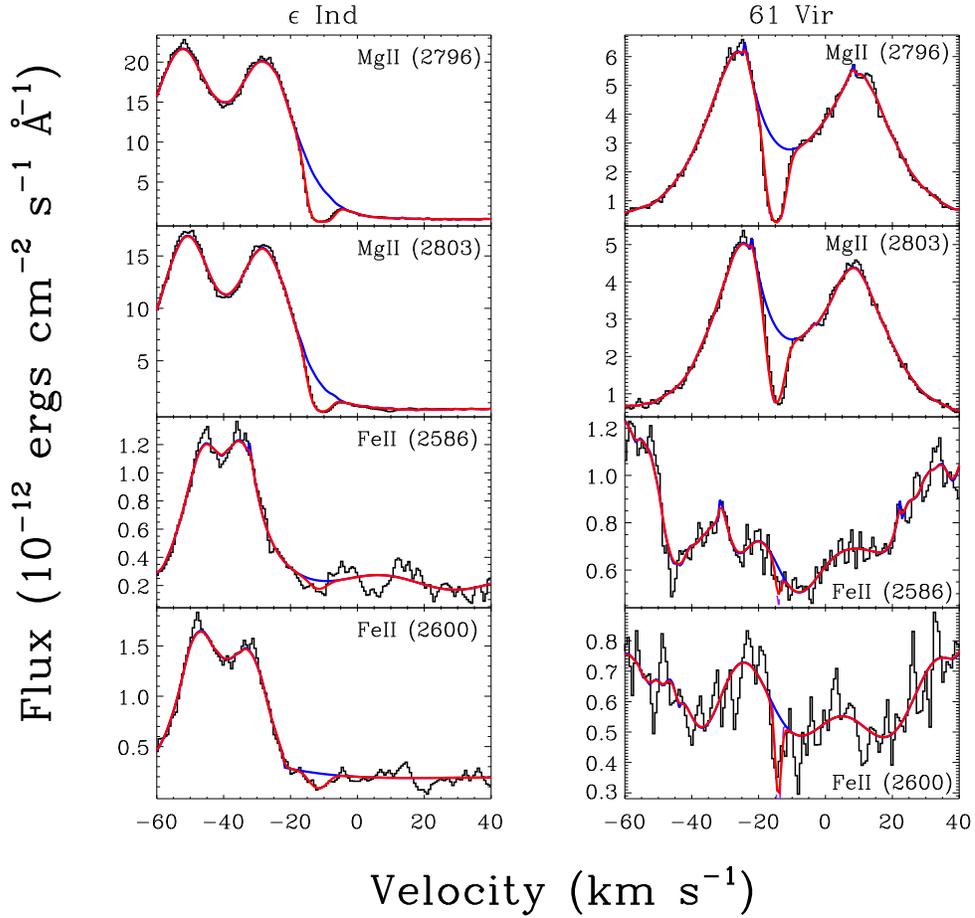}}
\caption{Simultaneous fits to the \ion{Mg}{2} and \ion{Fe}{2} lines displayed on a heliocentric velocity scale. The solid black histogram is the observed spectral data.  The solid blue lines are the fits to the stellar emission or continuum. The dashed lines (visible in multi-component fits) are profiles of each component. The solid red curves are the superposition of all components. If no ISM absorption features were detected, the spectra are not displayed.  \ion{Mn}{2} absorption is detected only in four of the most distant targets.  \label{vp1}}
\end{figure}

\begin{figure}
\centering
\figurenum{2b}
\centerline{\includegraphics{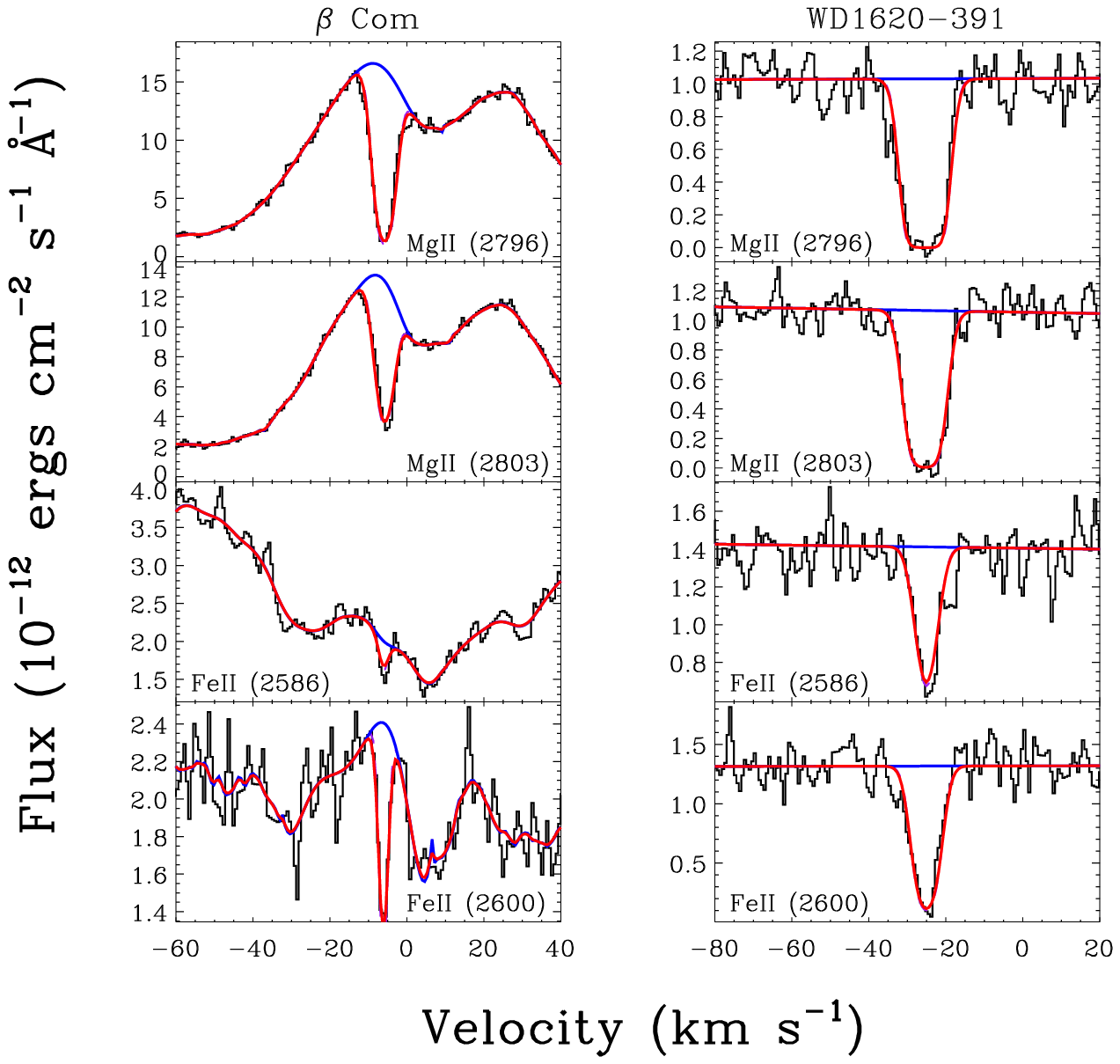}}
\caption{See caption of Figure \ref{vp1}. \label{vp2}}
\end{figure}

\begin{figure}
\centering
\figurenum{2c}
\centerline{\includegraphics{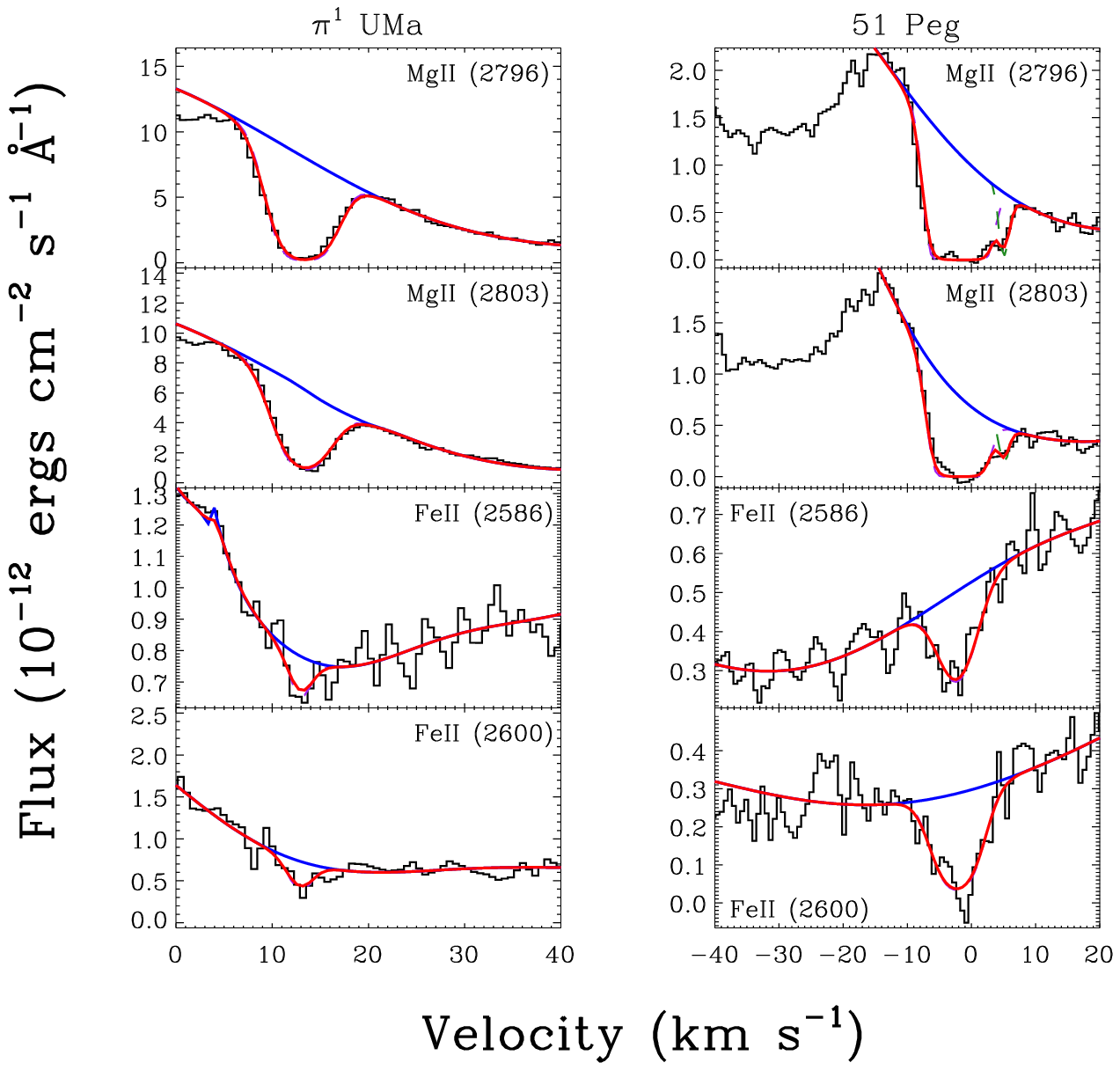}}
\caption{ See caption of Figure \ref{vp1}. \label{vp3}}
\end{figure}

\begin{figure}
\centering
\figurenum{2d}
\centerline{\includegraphics{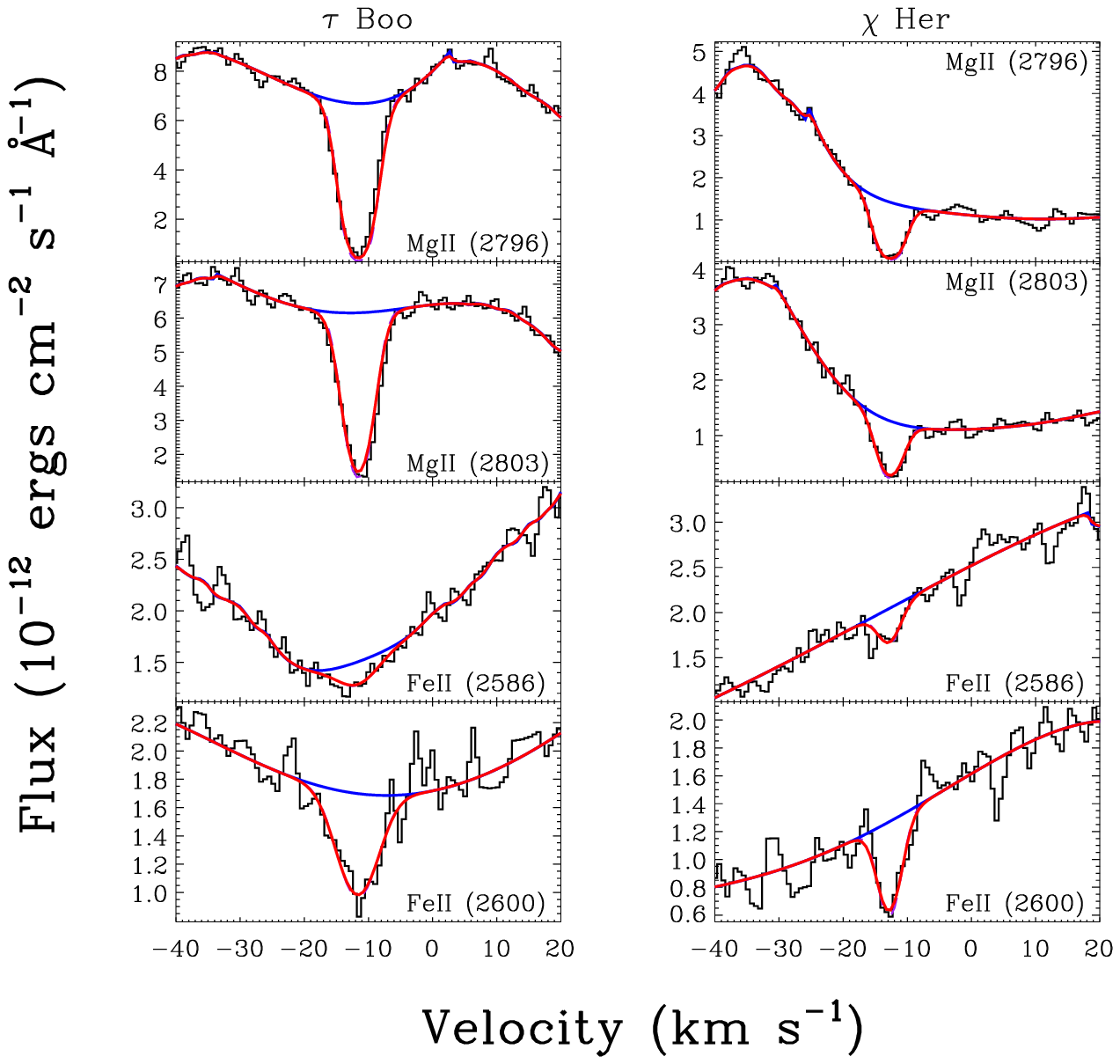}}
\caption{See caption of Figure \ref{vp1}.  \label{vp4}}
\end{figure}

\begin{figure}
\centering
\figurenum{2e}
\centerline{\includegraphics{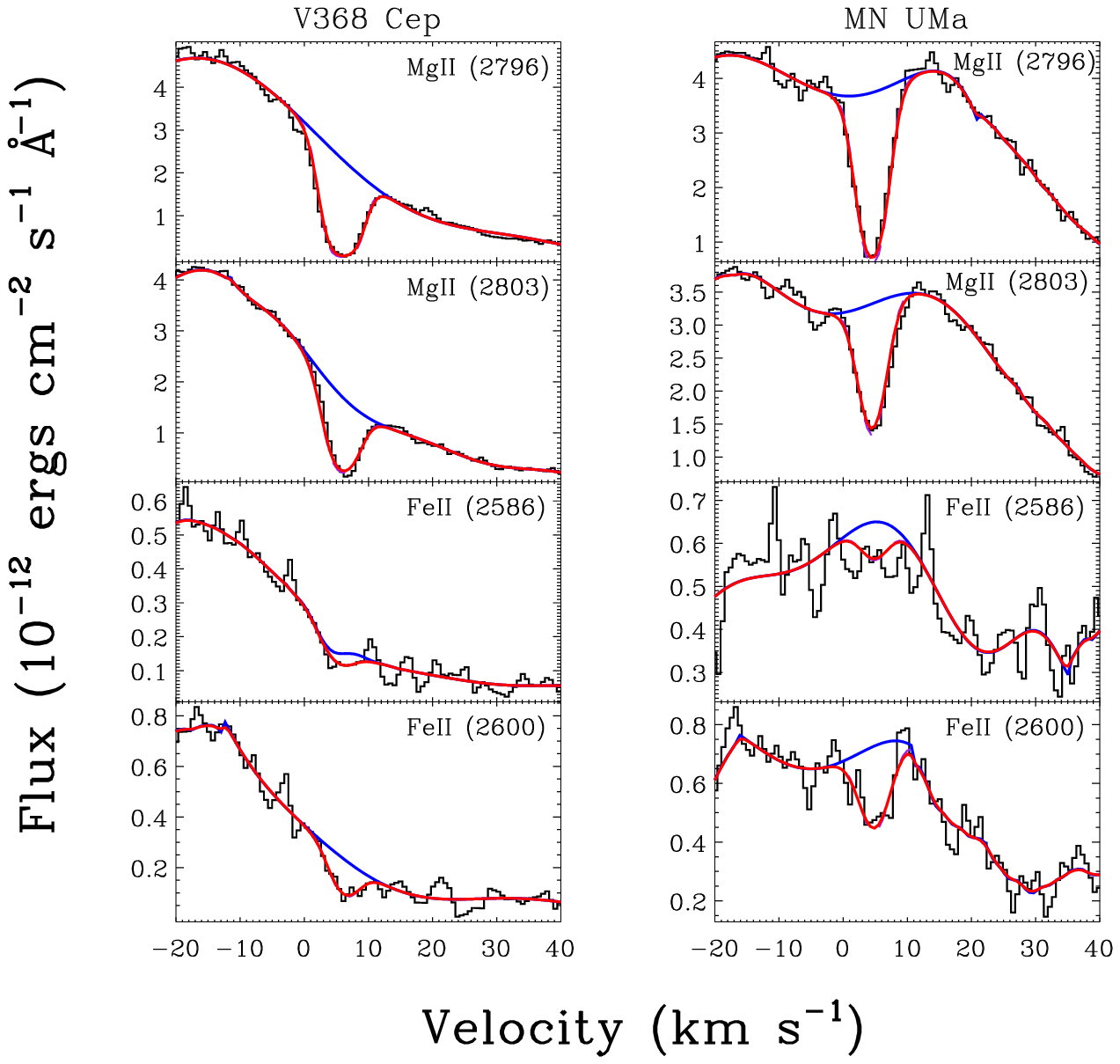}}
\caption{See caption of Figure \ref{vp1}.  \label{vp5}}
\end{figure}

\begin{figure}
\centering
\figurenum{2f}
\centerline{\includegraphics{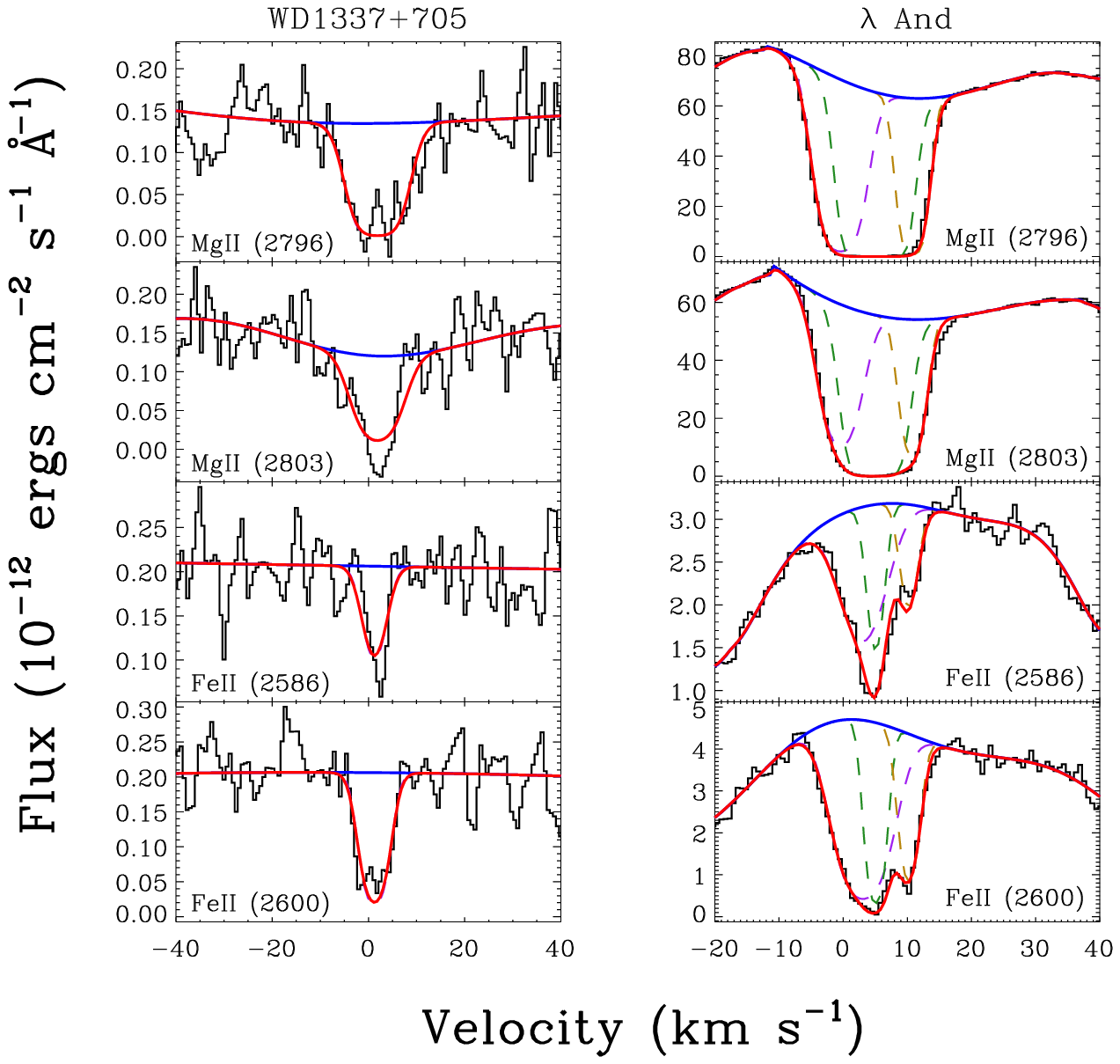}}
\caption{See caption of Figure \ref{vp1}.  \label{vp6}}
\end{figure}

\begin{figure}
\centering
\figurenum{2g}
\centerline{\includegraphics{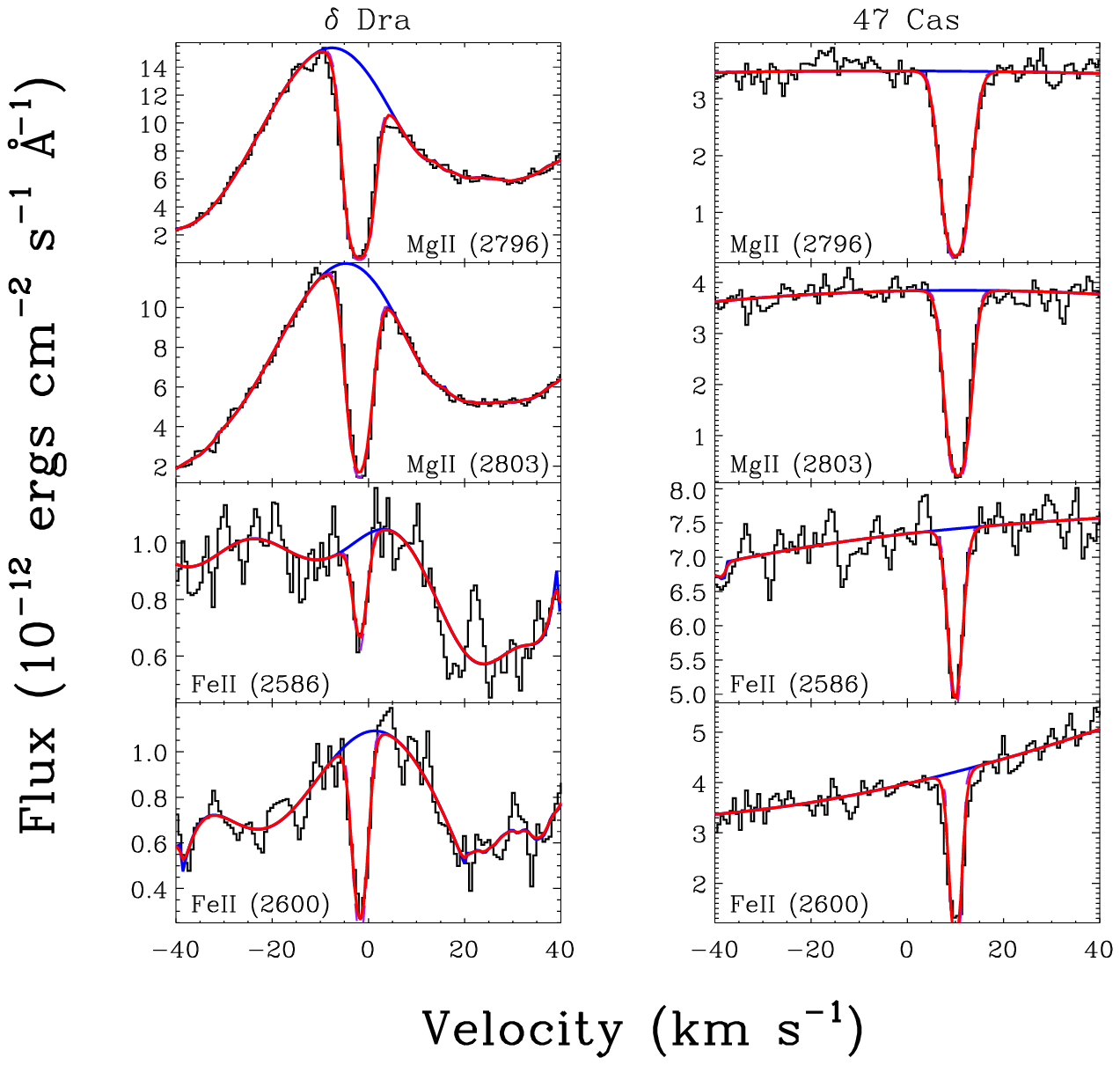}}
\caption{ See caption of Figure \ref{vp1}. \label{vp7}}
\end{figure}

\begin{figure}
\centering
\figurenum{2h}
\centerline{\includegraphics{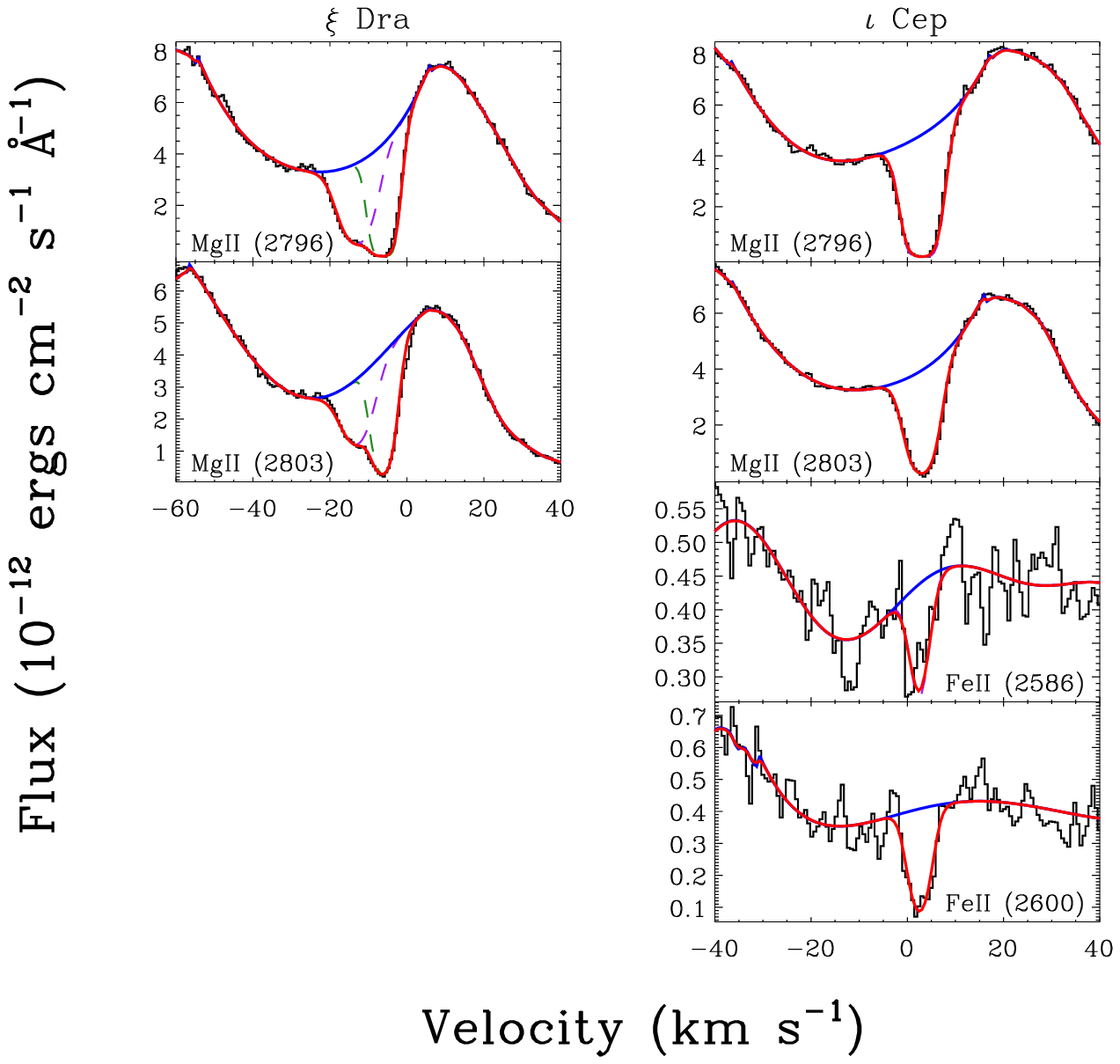}}
\caption{ See caption of Figure \ref{vp1}. \label{vp8}}
\end{figure}

\begin{figure}
\centering
\figurenum{2i}
\centerline{\includegraphics{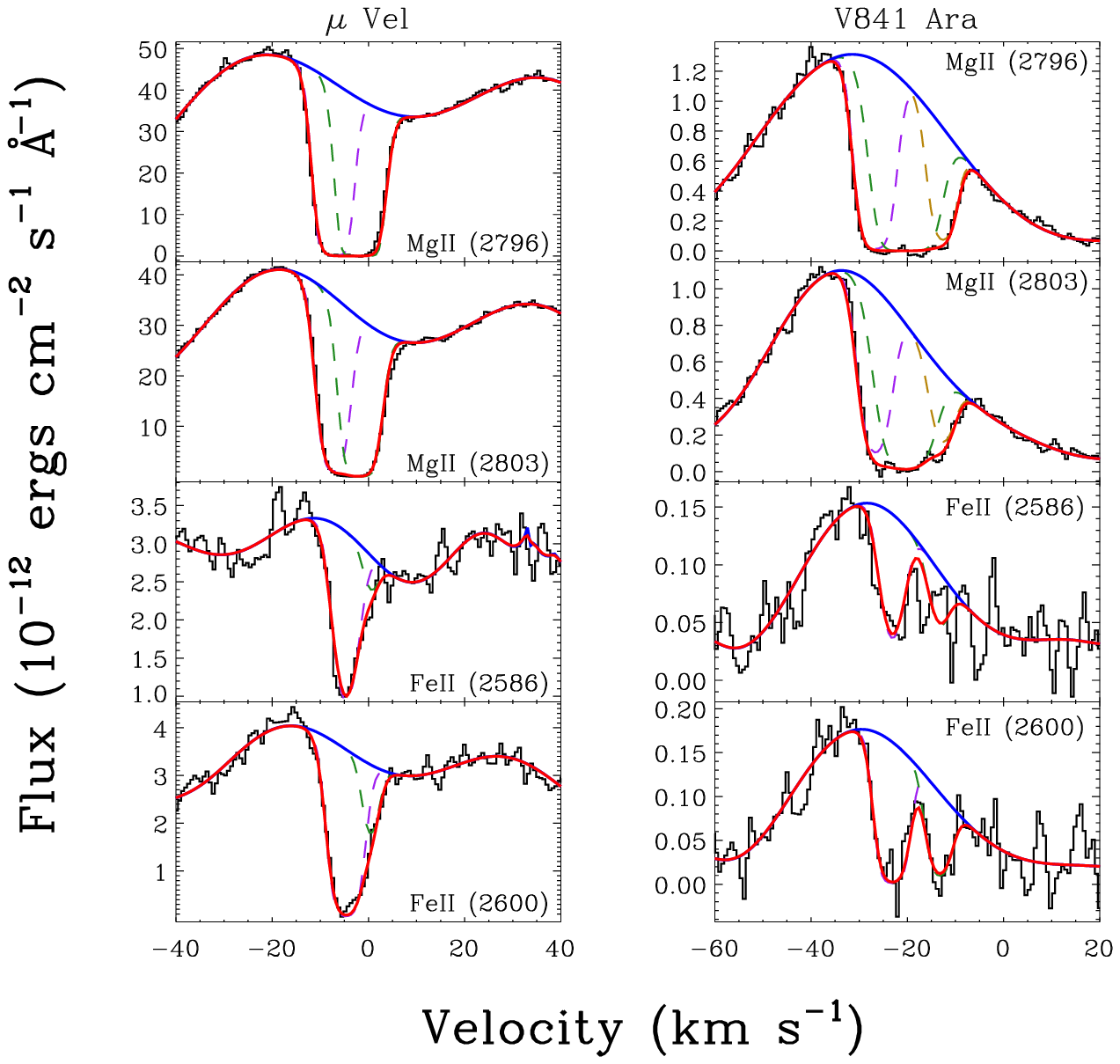}}
\caption{ See caption of Figure \ref{vp1}. \label{vp9}}
\end{figure}

\begin{figure}
\centering
\figurenum{2j}
\centerline{\includegraphics{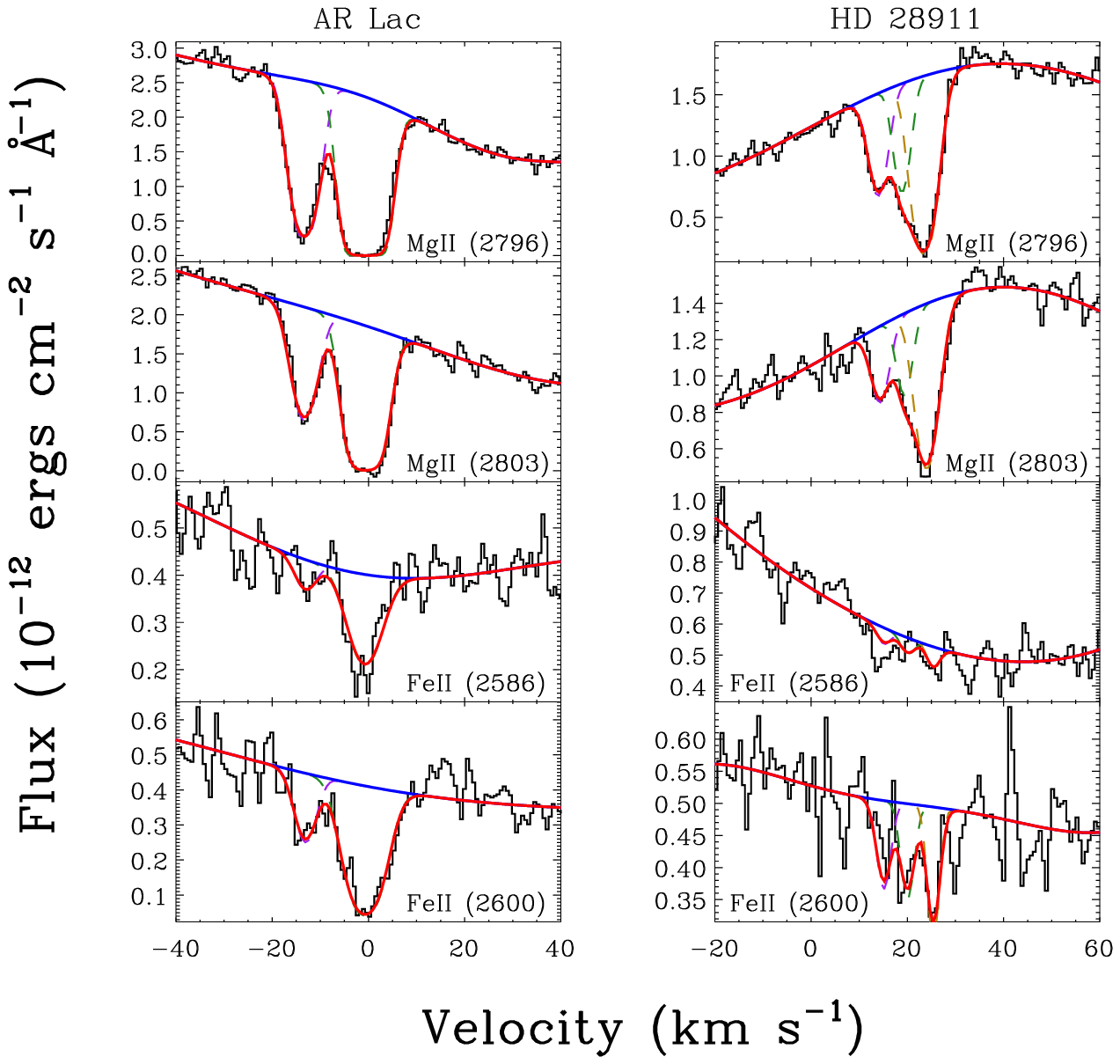}}
\caption{ See caption of Figure \ref{vp1}. \label{vp10}}
\end{figure}

\begin{figure}
\centering
\figurenum{2k}
\centerline{\includegraphics{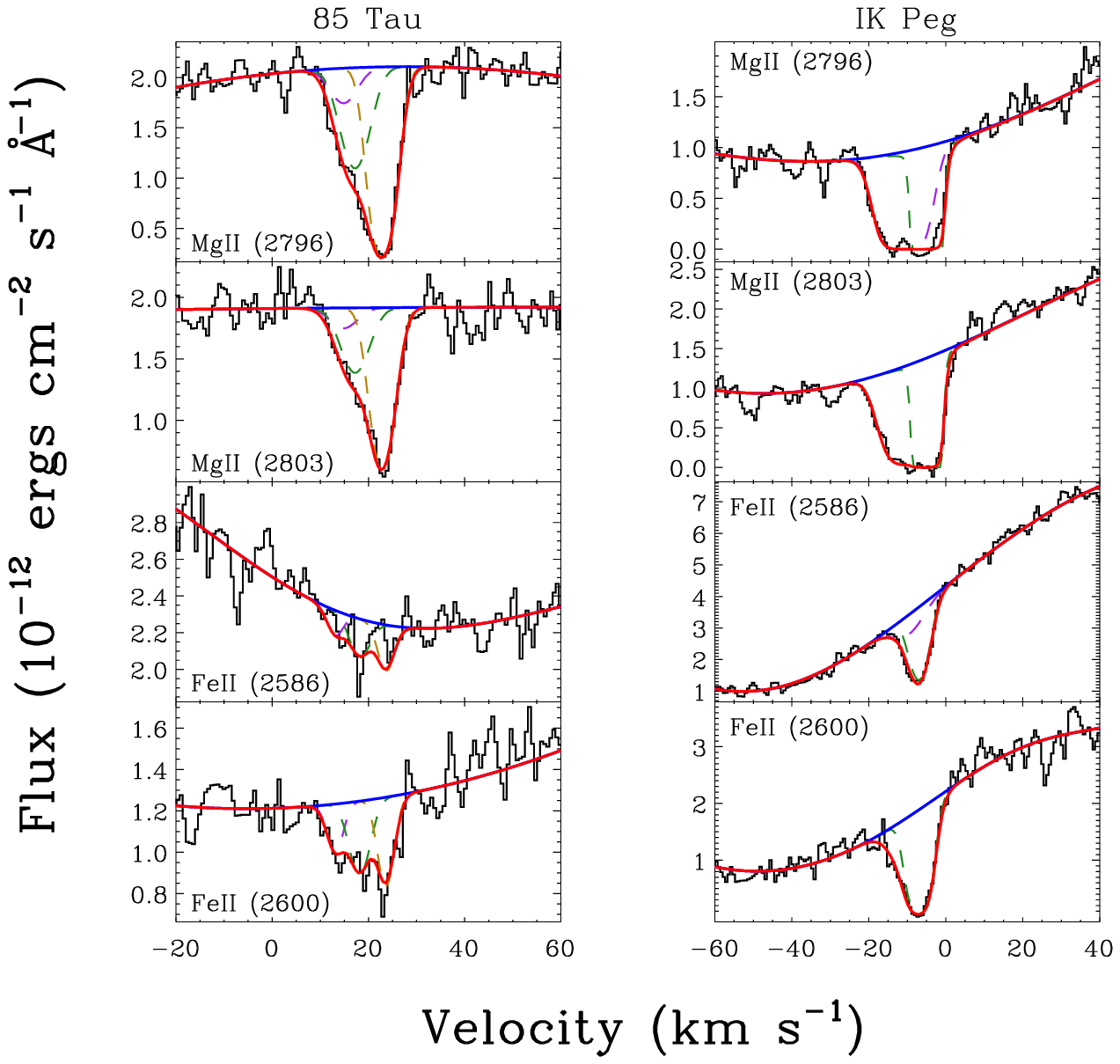}}
\caption{See caption of Figure \ref{vp1}.  \label{vp11}}
\end{figure}

\begin{figure}
\centering
\figurenum{2l}
\centerline{\includegraphics{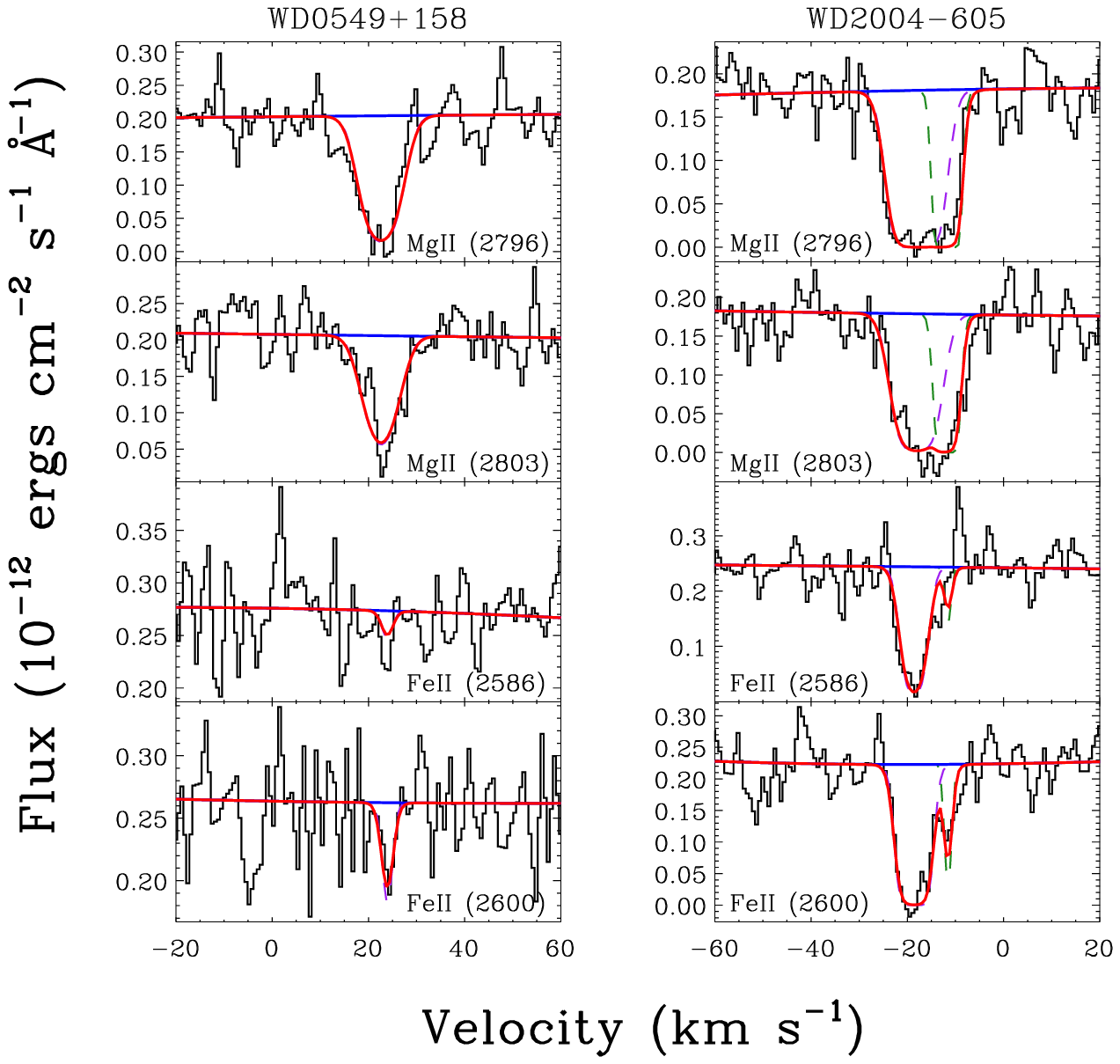}}
\caption{See caption of Figure \ref{vp1}.  \label{vp12}}
\end{figure}

\begin{figure}
\centering
\figurenum{2m}
\centerline{\includegraphics{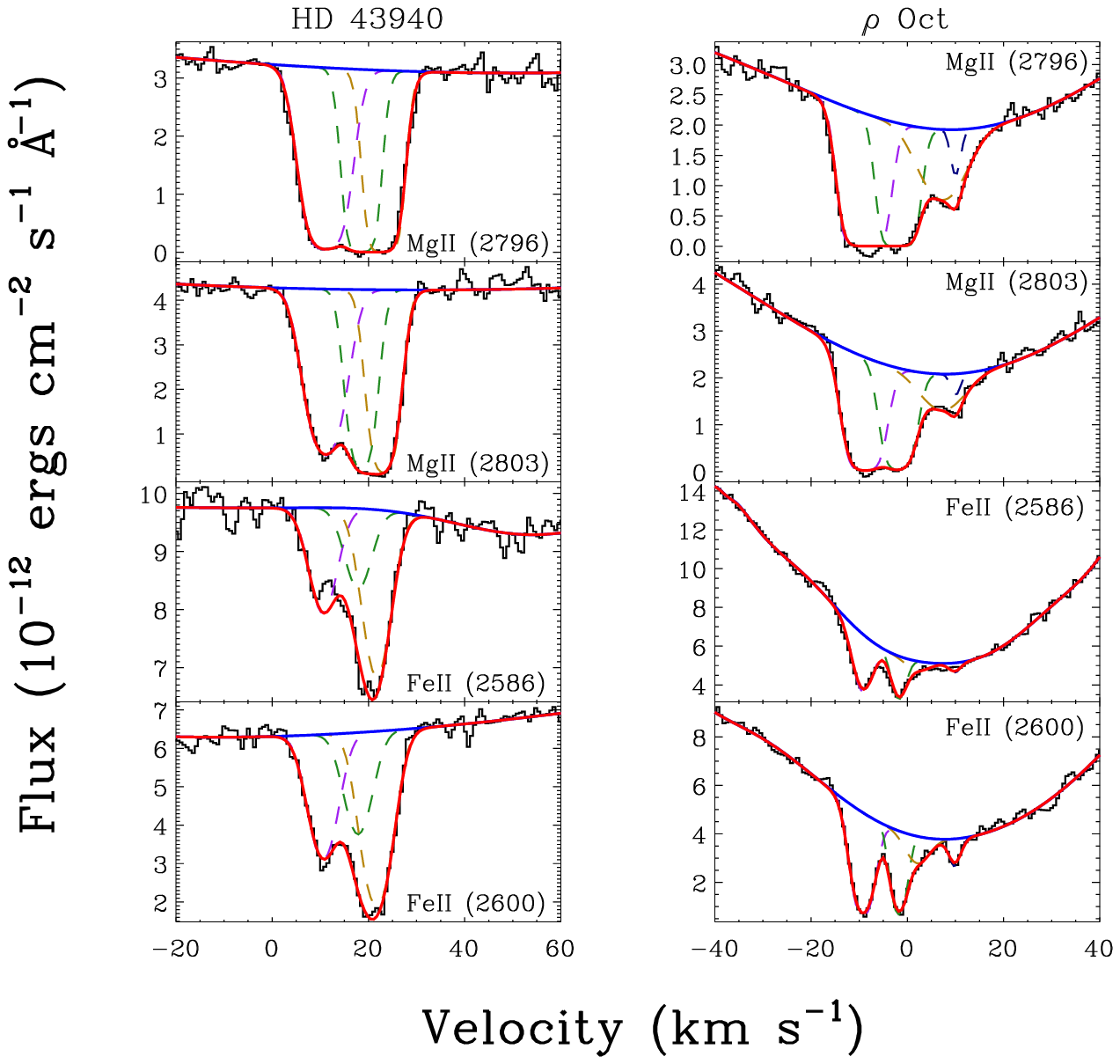}}
\caption{ See caption of Figure \ref{vp1}. \label{vp13}}
\end{figure}

\begin{figure}
\centering
\figurenum{2n}
\centerline{\includegraphics{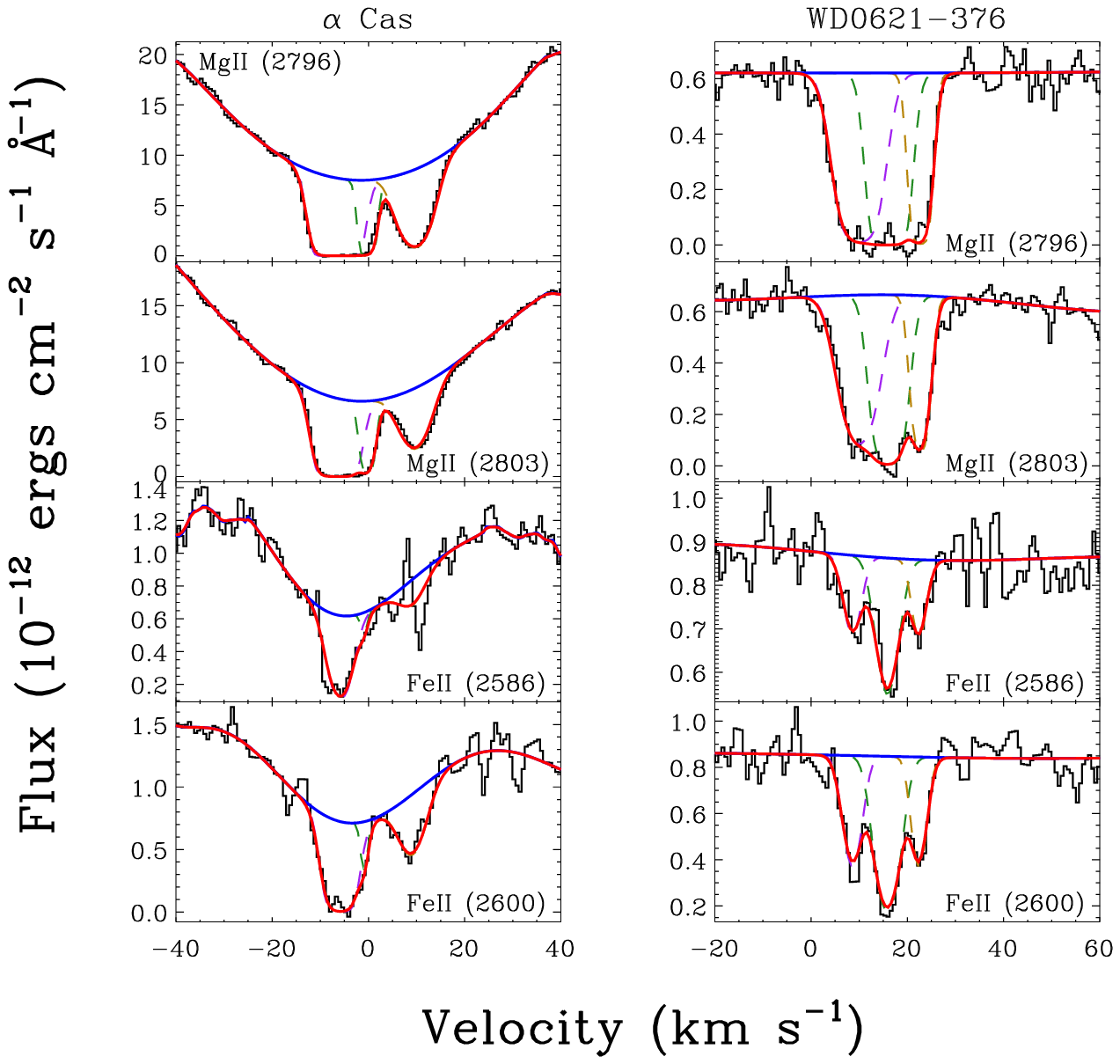}}
\caption{ See caption of Figure \ref{vp1}. \label{vp14}}
\end{figure}

\begin{figure}
\centering
\figurenum{2o}
\centerline{\includegraphics{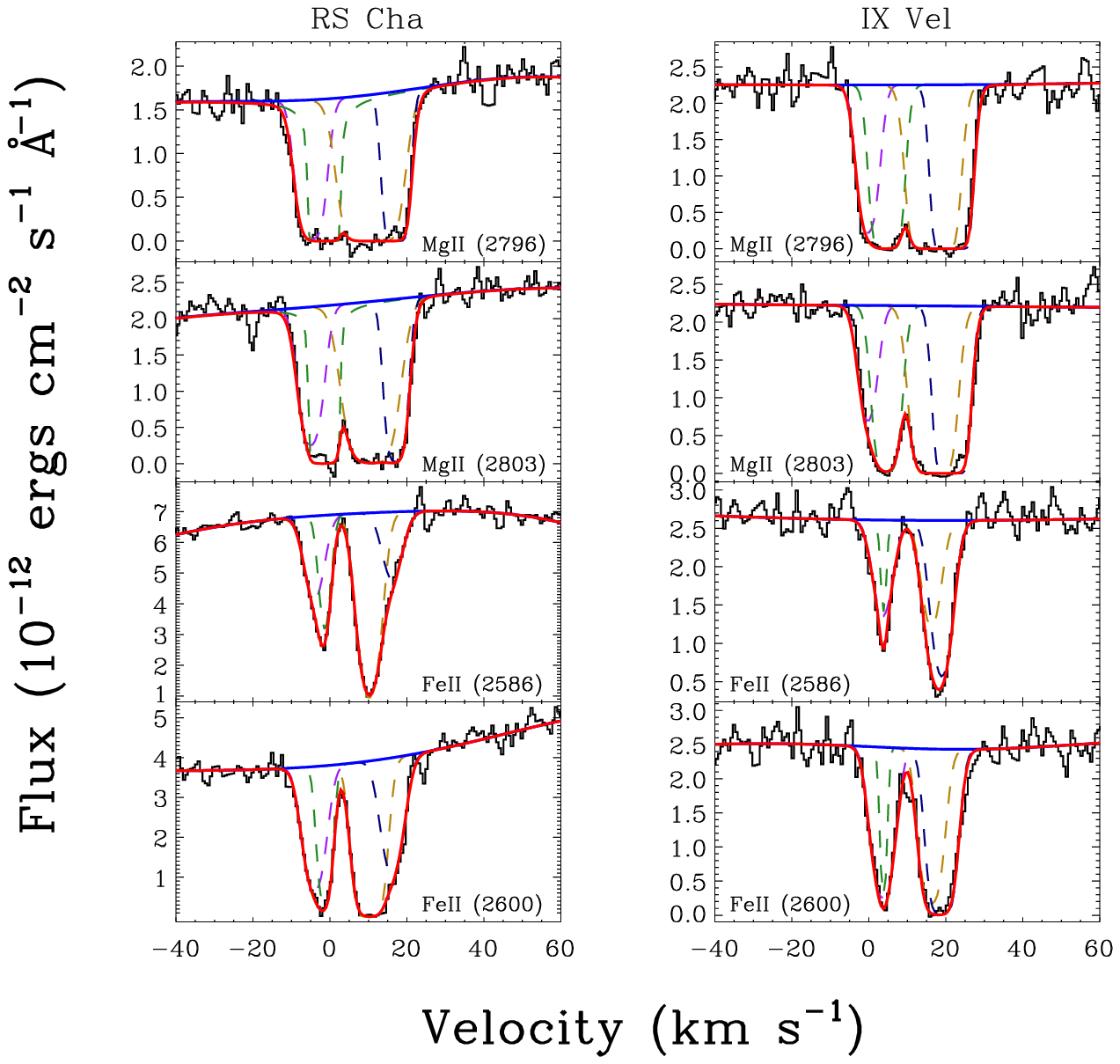}}
\caption{See caption of Figure \ref{vp1}.  \label{vp15}}
\end{figure}

\begin{figure}
\centering
\figurenum{2p}
\centerline{\includegraphics{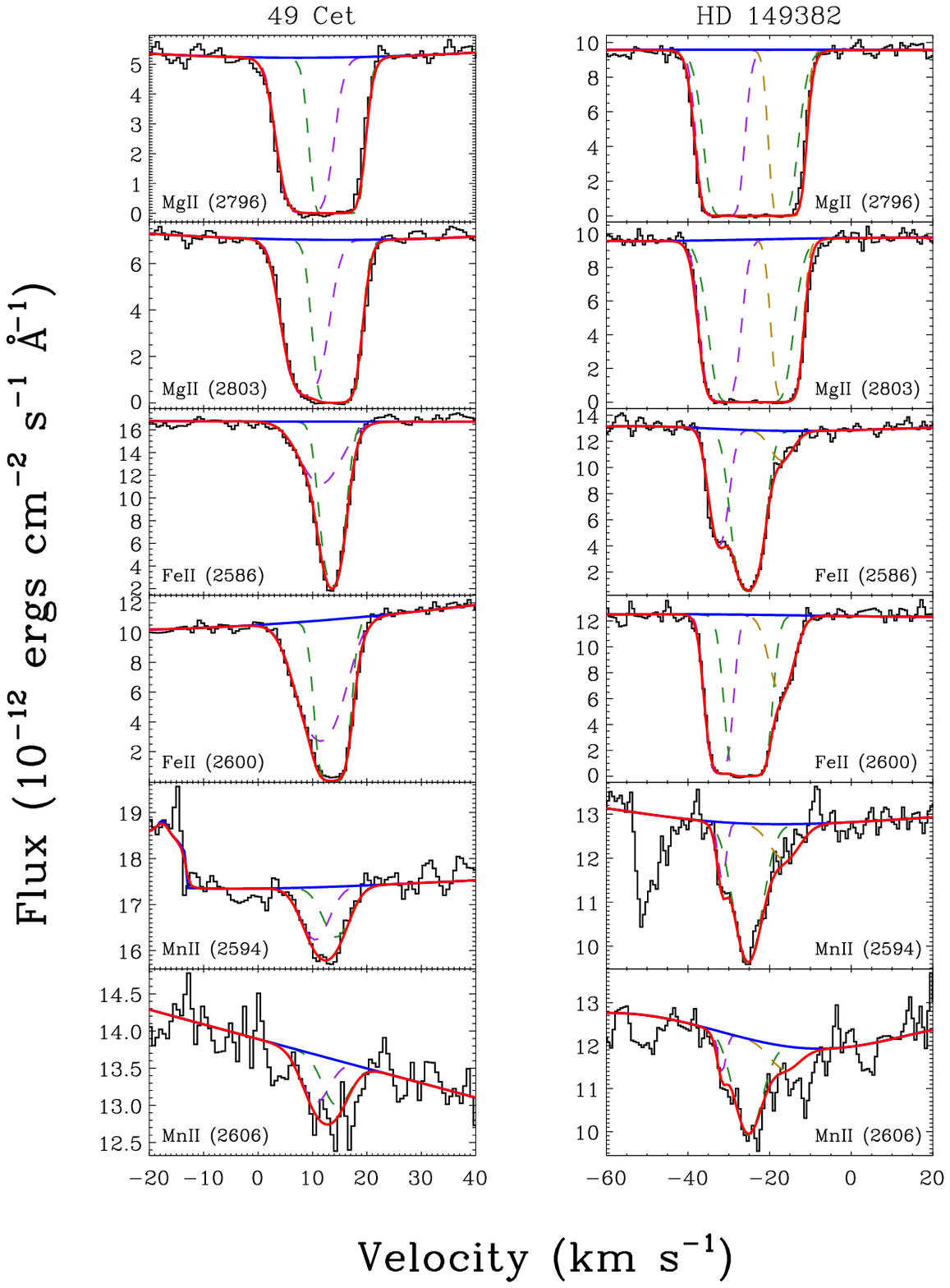}}
\caption{See caption of Figure \ref{vp1}. \label{vp16}}
\end{figure}

\begin{figure}
\centering
\figurenum{2q}
\centerline{\includegraphics{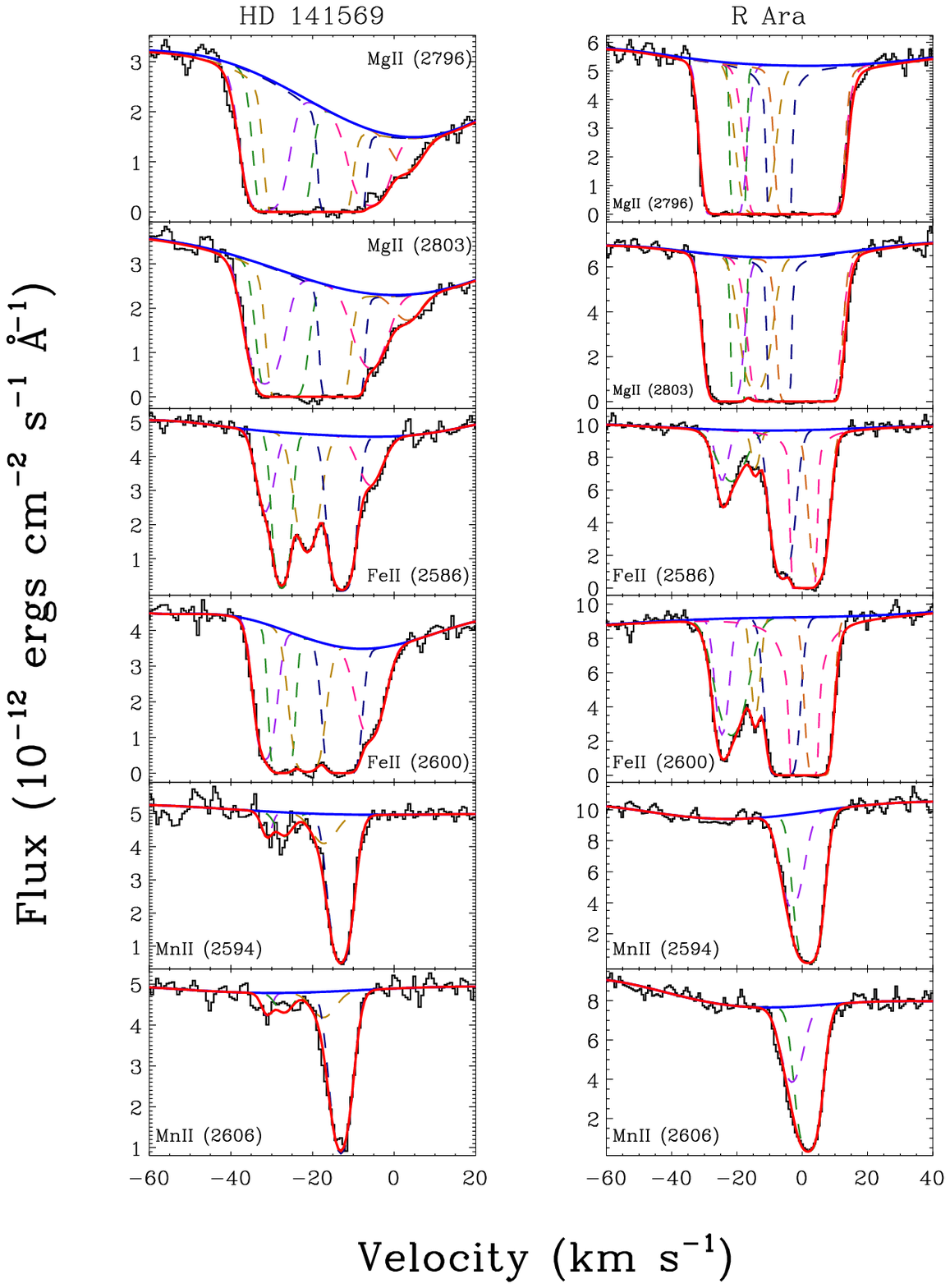}}
\caption{See caption of Figure \ref{vp1}.  \label{vp17}}
\end{figure}

	We measured interstellar absorption components in 34 lines of sight, with an average of 2.3 components per sight line. Every sight line contains \ion{Mg}{2} absorption, 33 show \ion{Fe}{2} absorption, and four have \ion{Mn}{2} absorption. In cases where \ion{Mg}{2} absorption was detected, but not a corresponding component in \ion{Fe}{2} or \ion{Mn}{2}, we estimated a 3$\sigma$ upper limit for the column density of a noise-obscured component. The fits are displayed in Figures \ref{vp1}--\ref{vp17}, and the final parameters and 3$\sigma$ upper limits are listed in Tables \ref{tabmg}--\ref{tabmn}.

\subsection{Results}

	Many of the longest sight lines in the SNAP sample have both saturated and blended components. In order to obtain an accurate fit to the LISM absorption profile, we often must assume in these cases that \ion{Mg}{2}, \ion{Fe}{2}, and \ion{Mn}{2} are well mixed and in thermodynamic equilibrium within a cloud.  We then constrain the spacing of the components in the saturated line using the velocities measured in the unsaturated lines. Such sight lines proved challenging because they provided minimal constraints on the Doppler parameter or column density of the components. For example, six ISM components were detected toward HD149730, a sight line 124 pc long (Figure \ref{vp17}). The velocity structure would be impossible to disentangle using \ion{Mg}{2} alone, because it is completely saturated and blended. Instead, the velocities determined from unsaturated \ion{Fe}{2} and \ion{Mn}{2} lines were used to fix the spacing of the six components in the \ion{Mg}{2} line, leaving the other parameters -- absolute velocities, Doppler widths, and column densities -- free in the modeling.
	
	In a few cases, the simultaneous fits indicated an apparent systematic error in the radial velocity measurements between the doublet components of the same ion. An example is $\beta$ Com (Figure \ref{vp2}). Because such anomalies affect only a few of the spectra, the cause is unlikely to be due to the StarCAT wavelength distortion correction.  It could be associated with the complexity of the continuum placement in certain cases, or perhaps partial saturation in the stronger member of the doublet.  In any event, the discrepancy is only of order 0.4  km~s$^{-1}$, comparable to the internal precision of the StarCAT processing in the most favorable (i.e., high S/N) cases (0.3 km~s$^{-1}$; \citealt{ayres10}), and inconsequential for the measured absorption parameters.


\subsection{Final Fit Parameters}\label{sec:tables}

Tables \ref{tabmg}--\ref{tabmn} list the final parameters of the fitting procedure: heliocentric radial velocities ($v$), Doppler parameters ($b$), and $\log$ column densities ($\log N$). Each value is a weighted mean based on the individual and simultaneous fits. Components seen in \ion{Mg}{2} but not in \ion{Fe}{2} or \ion{Mn}{2} have 3$\sigma$ upper limits for the latter listed in the column density columns.

\section{Analysis}
\label{analysis.s}

Among the 34 sight lines analyzed, we detected 76 absorption components in \ion{Mg}{2}, 71 in \ion{Fe}{2}, and 11 in \ion{Mn}{2}. In parts of this discussion, \ion{Mn}{2} is excluded due to the small number of detected components.

Figure \ref{fighist} shows the distribution of derived parameters for all observed ions, along with the complete NUV LISM dataset \citep{redfield02,redfield04sw}. Values are listed in Tables \ref{tabmg}--\ref{tabmn}. Outliers (e.g., high Doppler widths) can result from saturated and blended lines for which the fits are poorly constrained. The more common occurrence of saturated and blended lines at larger distances demonstrates the unique advantages of studying the ISM locally.



\begin{deluxetable}{llcccc}
\tablewidth{0pt}
\tabletypesize{\tiny}
\tablecaption{Fit Parameters for New \ion{Mg}{2} LISM Components \label{tabmg}}
\tablehead{&& Component & $v$\tablenotemark{a} & $b$ & $\log N_{\rm{MgII}}$ \\
HD No. & Other Name & No. & (km s$^{-1}$) & (km s$^{-1}$) & log(cm$^{-2})$ \\}
\startdata
209100	&	 $\epsilon$ Ind 		&	1	&	$-$10.83	$\pm$	0.35		&	3.099	$\pm$	0.031	&	12.846	$\pm$	0.054	\\
115617	&	 61 Vir 			&	1	&	$-$14.74	$\pm$	0.42		&	2.69		$\pm$	0.11		&	12.4710	$\pm$	0.0089	\\
114710	&	 $\beta$ Com		&	1	&	$-$5.83	$\pm$	0.29		&	2.923	$\pm$	0.035	&	12.492	$\pm$	0.014	\\
       		&	 WD1620--391		&	1	&	$-$25.42	$\pm$	0.37		&	4.29		$\pm$	0.52		&	13.11	$\pm$	0.18		\\
72905	&	 $\pi^1$ UMa		&	1	&	13.29	$\pm$	0.24		&	3.04		$\pm$	0.11		&	12.661	$\pm$	0.032	\\
217014	&	 51 Peg			&	1	&	$-$1.94	$\pm$	0.31		&	3.09		$\pm$	0.17		&	13.38	$\pm$	0.13		\\
		&					&	2	&	5.01		$\pm$	0.55		&	0.93		$\pm$	0.52		&	11.94	$\pm$	0.19		\\
120136	&	 $\tau$ Boo 		&	1	&	$-$11.61	$\pm$	0.26		&	2.74		$\pm$	0.13		&	12.513	$\pm$	0.055	\\
142373	&	 $\chi$ Her  		&	1	&	$-$12.69	$\pm$	0.16		&	2.21		$\pm$	0.12		&	12.465	$\pm$	0.030	\\
220140	&	 V368 Cep    	   	&	1	&	6.04		$\pm$	0.23		&	2.76		$\pm$	0.32		&	12.648	$\pm$	0.023	\\
97334	&	 MN UMa         		&	1	&	4.54		$\pm$	0.19		&	2.636	$\pm$	0.086	&	12.270	$\pm$	0.030	\\
       		&	 WD1337+705    	&	1	&	1.83		$\pm$	0.39		&	4.7		$\pm$	1.2		&	12.95	$\pm$	0.23		\\
222107	&	 $\lambda$ And  	&	1	&	0.14		$\pm$	0.74		&	3.62		$\pm$	0.28		&	12.75	$\pm$	0.16		\\
		&					&	2	&	4.97		$\pm$	0.30		&	3.4		$\pm$	1.3		&	13.17	$\pm$	0.22		\\
		&					&	3	&	10.17	$\pm$	0.85		&	2.58		$\pm$	0.51		&	12.66	$\pm$	0.16		\\
180711	&	 $\delta$ Dra   		&	1	&	$-$1.90	$\pm$	0.28		&	2.715	$\pm$	0.073	&	12.655	$\pm$	0.024	\\
12230	&	 47 Cas         		&	1	&	10.27	$\pm$	0.17		&	2.49		$\pm$	0.26		&	12.62	$\pm$	0.13		\\
163588	&	$\xi$ Dra			&	1	&	$-$13.72	$\pm$	0.76		&	4.57		$\pm$	0.50		&	12.505	$\pm$	0.081	\\
		&					&	2	&	$-$6.27	$\pm$	0.42		&	3.49		$\pm$	0.26		&	12.852	$\pm$	0.030	\\
216228	&	$\iota$ Cep		&	1	&	3.06		$\pm$	0.20		&	3.407	$\pm$	0.055	&	12.880	$\pm$	0.031	\\
93497	&	$\mu$ Vel			&	1	&	$-$7.2	$\pm$	1.2		&	3.08		$\pm$	0.40		&	12.91	$\pm$	0.22		\\
		&					&	2	&	1.2		$\pm$	1.7		&	3.33		$\pm$	0.21		&	13.13	$\pm$	0.16		\\
149499	&	V841 Ara			&	1	&	$-$25.90	$\pm$	0.63		&	3.40		$\pm$	0.21		&	12.87	$\pm$	0.27		\\
		&					&	2	&	$-$19.56	$\pm$	0.91		&	2.5		$\pm$	2.3		&	13.22	$\pm$	0.80		\\
		&					&	3	&	$-$13.32	$\pm$	0.68		&	3.13		$\pm$	0.45		&	12.58	$\pm$	0.17		\\
210334	&	AR Lac		&	1	&	$-$13.32	$\pm$	0.34		&	3.19		$\pm$	0.11		&	12.478	$\pm$	0.014	\\
		&					&	2	&	$-$0.63	$\pm$	0.33		&	3.82		$\pm$	0.43		&	13.08	$\pm$	0.17		\\
28911	&	 HIP21267       		&	1	&	14.30	$\pm$	0.26		&	2.47		$\pm$	0.41		&	11.894	$\pm$	0.065	\\
		&					&	2	&	20.3		$\pm$	2.2		&	2.4		$\pm$	1.6		&	11.84	$\pm$	0.38		\\
		&					&	3	&	23.83	$\pm$	0.45		&	3.07		$\pm$	0.87		&	12.17	$\pm$	0.33		\\
28677	&	 85 Tau         		&	1	&	13.81	$\pm$	0.77		&	2.7		$\pm$	1.2		&	11.42	$\pm$	0.25		\\
		&					&	2	&	18.6		$\pm$	1.1		&	3.42		$\pm$	0.81		&	12.09	$\pm$	0.12		\\
		&					&	3	&	23.26	$\pm$	0.36		&	2.91		$\pm$	0.32		&	12.409	$\pm$	0.055	\\
204188	&	IK Peg			&	1	&	$-$12.2	$\pm$	1.7		&	5.22		$\pm$	0.92		&	12.92	$\pm$	0.21		\\
		&					&	2	&	$-$5.33	$\pm$	0.61		&	2.51		$\pm$	0.83		&	13.12	$\pm$	0.65		\\
		&	WD0549+158		&	1	&	22.58	$\pm$	0.81		&	4.09		$\pm$	0.32		&	12.631	$\pm$	0.044	\\
		&	WD2004--605		&	1	&	$-$17.9	$\pm$	1.7		&	3.89		$\pm$	0.54		&	13.17	$\pm$	0.33		\\
		&					&	2	&	$-$12.00	$\pm$	0.90		&	2.2		$\pm$	1.4		&	12.89	$\pm$	0.32		\\
9672		&	 49 Cet         		&	1	&	9.0		$\pm$	1.3		&	3.88		$\pm$	0.51		&	12.97	$\pm$	0.17		\\
		&					&	2	&	14.4		$\pm$	1.1		&	2.98		$\pm$	0.42		&	13.36	$\pm$	0.18		\\
43940	&	HR2265			&	1	&	11.15	$\pm$	0.24		&	4.55		$\pm$	0.20		&	12.878	$\pm$	0.030	\\
		&					&	2	&	18.26	$\pm$	0.72		&	2.68		$\pm$	0.70		&	12.52	$\pm$	0.30		\\*
		&					&	3	&	23.18	$\pm$	0.86		&	3.24		$\pm$	0.35		&	12.83	$\pm$	0.22		\\*
137333	&	$\rho$ Oct			&	1	&	$-$8.82	$\pm$	0.59		&	3.69		$\pm$	0.16		&	13.16	$\pm$	0.10		\\*
		&					&	2	&	$-$1.9	$\pm$	1.0		&	2.91		$\pm$	0.47		&	12.99	$\pm$	0.28		\\*
		&					&	3	&	3.2		$\pm$	3.4		&	4.4		$\pm$	1.3		&	12.33	$\pm$	0.16		\\*
		&					&	4	&	9.61		$\pm$	0.57		&	3.3		$\pm$	1.3		&	11.68	$\pm$	0.45		\\*
3712		&	$\alpha$ Cas		&	1	&	$-$6.75	$\pm$	0.45\tablenotemark{b}		&	3.57		$\pm$	0.16		&	13.29	$\pm$	0.20		\\*
		&					&	2	&	$-$2.46						&	2.83		$\pm$	0.22		&	12.91	$\pm$	0.20		\\*
		&					&	3	&	9.74		$\pm$	0.50		&	3.880	$\pm$	0.073	&	12.5526	$\pm$	0.0058	\\*
149382	&	HIP81145			&	1	&	$-$32.02	$\pm$	0.55\tablenotemark{b}		&	3.68		$\pm$	0.16		&	13.21	$\pm$	0.12		\\*
		&					&	2	&	$-$24.57					&	4.9		$\pm$	1.3		&	13.92	$\pm$	0.48		\\*
		&					&	3	&	$-$15.70					&	2.89		$\pm$	0.23		&	13.059	$\pm$	0.099	\\*
		&	WD0621--376		&	1	&	10.31	$\pm$	0.65		&	4.68		$\pm$	0.41		&	12.81	$\pm$	0.14		\\*
		&					&	2	&	16.22	$\pm$	0.54		&	3.2		$\pm$	1.6		&	13.11	$\pm$	0.26		\\*
		&					&	3	&	22.3		$\pm$	1.4		&	2.50		$\pm$	0.81		&	12.667	$\pm$	0.077	\\*
75747	&	RS Cha			&	1	&	$-$4.86	$\pm$	0.12\tablenotemark{b}		&	3.31		$\pm$	0.47		&	12.74	$\pm$	0.14		\\*
		&					&	2	&	$-$1.37						&	2.12		$\pm$	0.98		&	13.46	$\pm$	0.69		\\*
		&					&	3	&	10.68					&	4.7		$\pm$	1.6		&	13.40	$\pm$	0.81		\\*
		&					&	4	&	17.27					&	2.78		$\pm$	0.26		&	12.84	$\pm$	0.15		\\*
		&	IX Vel			&	1	&	1.2		$\pm$	1.9		&	3.22		$\pm$	0.52		&	12.69	$\pm$	0.32		\\*
		&					&	2	&	4.91		$\pm$	0.40		&	2.72		$\pm$	0.41		&	12.88	$\pm$	0.27		\\*
		&					&	3	&	16.44	$\pm$	0.49		&	3.91		$\pm$	0.29		&	13.56	$\pm$	0.19		\\
		&					&	4	&	20.80	$\pm$	0.69		&	3.49		$\pm$	0.64		&	13.30	$\pm$	0.24		\\
141569	&	HIP77542			&	1	&	$-$31.24	$\pm$	0.54\tablenotemark{b}		&	4.44		$\pm$	0.41		&	13.06	$\pm$	0.13		\\
		&					&	2	&	$-$26.90					&	3.0		$\pm$	1.2		&	13.2		$\pm$	1.2		\\
		&					&	3	&	$-$20.54					&	3.8		$\pm$	1.2		&	13.62	$\pm$	0.86		\\
		&					&	4	&	$-$12.24					&	2.45		$\pm$	0.48		&	13.60	$\pm$	0.65		\\
		&					&	5	&	$-$5.36						&	4.85		$\pm$	0.45		&	12.54	$\pm$	0.12		\\
		&					&	6	&	2.84						&	4.18		$\pm$	0.98		&	11.974	$\pm$	0.090	\\
149730	&	R Ara			&	1	&	$-$24.17	$\pm$	0.59\tablenotemark{b}		&	4.32		$\pm$	0.34		&	13.75	$\pm$	0.14		\\
		&					&	2	&	$-$19.59					&	1.29		$\pm$	0.48		&	13.18	$\pm$	0.21		\\
		&					&	3	&	$-$14.25					&	3.5		$\pm$	2.3		&	13.03	$\pm$	0.16		\\
		&					&	4	&	$-$7.05						&	4.1		$\pm$	2.4		&	14.16	$\pm$	0.42		\\
		&					&	5	&	$-$2.39						&	6.4		$\pm$	1.9		&	14.52	$\pm$	0.35		\\
		&					&	6	&	2.24						&	4.31		$\pm$	0.64		&	14.84	$\pm$	0.28		\\
\enddata
\tablenotetext{a}{heliocentric radial velocity}
\tablenotetext{b}{This uncertainty applies to all other component velocities of the sight line with no uncertainty given. These components were assumed to have a fixed spacing when fitting velocities, based on unsaturated \ion{Fe}{2} or \ion{Mn}{2} lines.}
\end{deluxetable}


\begin{deluxetable} {llcccc}
\tablewidth{0pt}
\tabletypesize{\tiny}
\tablecaption{Fit Parameters for New \ion{Fe}{2} LISM Components \label{tabfe}}
\tablehead{& & Component & $v$\tablenotemark{a} & $b$ & $\log N_{\rm{FeII}}$ \\
HD No. & Other Name & Number & (km s$^{-1}$) & (km s$^{-1}$) & log(cm$^{-2})$\\}
\startdata
209100	&	 $\epsilon$ Ind 		&	1	&	$-$11.34	$\pm$	0.28		&	2.77	$\pm$	0.63	&	12.62	$\pm$	0.11		\\
115617	&	 61 Vir 			&	1	&	$-$14.24	$\pm$	0.52		&	1.20	$\pm$	0.65	&	11.96	$\pm$	0.10		\\
114710	&	 $\beta$ Com		&	1	&	$-$6.04	$\pm$	0.19		&	1.37	$\pm$	0.27	&	12.08	$\pm$	0.11		\\
       	&	 WD1620--391			&	1	&	$-$25.04	$\pm$	0.42		&	3.62	$\pm$	0.15	&	12.995	$\pm$	0.039	\\
72905	&	 $\pi^1$ UMa		&	1	&	13.10	$\pm$	0.34		&	1.24	$\pm$	0.48	&	12.029	$\pm$	0.089	\\
217014	&	 51 Peg			&	1	&	$-$2.21	$\pm$	0.28		&	3.76	$\pm$	0.44	&	12.942	$\pm$	0.046	\\
		&					&	2	&	...						&	...				&	$<$12.1					\\
120136	&	 $\tau$ Boo 		&	1	&	$-$11.68	$\pm$	0.36		&	4.08	$\pm$	0.51	&	12.398	$\pm$	0.051	\\
142373	&	 $\chi$ Her  		&	1	&	$-$12.76	$\pm$	0.43		&	2.20	$\pm$	0.40	&	12.281	$\pm$	0.065	\\
220140	&	 V368 Cep       		&	1	&	6.27		$\pm$	0.95		&	2.72	$\pm$	0.49	&	12.49	$\pm$	0.14		\\
97334	&	 MN UMa         		&	1	&	4.94		$\pm$	0.39		&	3.19	$\pm$	0.93	&	12.22	$\pm$	0.18		\\
       		&	 WD1337+705    	&	1	&	1.36		$\pm$	0.54		&	2.99	$\pm$	0.91	&	12.895	$\pm$	0.073	\\
222107	&	 $\lambda$ And  	&	1	&	1.1		$\pm$	2.2		&	4.0	$\pm$	1.2	&	12.42	$\pm$	0.41		\\
		&					&	2	&	4.52		$\pm$	0.41		&	2.58	$\pm$	0.79	&	12.92	$\pm$	0.26		\\
		&					&	3	&	10.28	$\pm$	0.22		&	1.82	$\pm$	0.22	&	12.533	$\pm$	0.062	\\
180711	&	 $\delta$ Dra   		&	1	&	$-$1.75	$\pm$	0.16		&	1.75	$\pm$	0.26	&	12.483	$\pm$	0.049	\\
12230	&	 47 Cas         		&	1	&	10.036	$\pm$	0.078	&	1.93	$\pm$	0.12	&	12.480	$\pm$	0.021	\\
163588	&	$\xi$ Dra			&	1,2	&	...						&	...				&			$<$12.3			\\
216228	&	$\iota$ Cep		&	1	&	2.56		$\pm$	0.34		&	2.66	$\pm$	0.27	&	12.684	$\pm$	0.042	\\
93497	&	$\mu$ Vel			&	1	&	$-$5.25	$\pm$	0.59		&	2.78	$\pm$	0.67	&	12.86	$\pm$	0.21		\\
		&					&	2	&	$-$1.3	$\pm$	2.0		&	2.83	$\pm$	0.91	&	12.22	$\pm$	0.45		\\
149499	&	V841 Ara			&	1	&	$-$23.18	$\pm$	0.27		&	3.03	$\pm$	0.49	&	13.192	$\pm$	0.092	\\
		&					&	2	&	...						&	...				&	$<$12.6					\\
		&					&	3	&	$-$13.3	$\pm$	1.3		&	2.20	$\pm$	0.53	&	12.83	$\pm$	0.18		\\
210334	&	AR Lac		&	1	&	$-$13.12	$\pm$	0.50		&	2.6	$\pm$	1.0	&	12.25	$\pm$	0.11		\\
		&					&	2	&	$-$0.77	$\pm$	0.37		&	4.42	$\pm$	0.86	&	13.040	$\pm$	0.032	\\
28911	&	 HIP21267       		&	1	&	14.35	$\pm$	0.81		&	1.60	$\pm$	0.76	&	11.69	$\pm$	0.32		\\
		&					&	2	&	18.9		$\pm$	1.6		&	1.91	$\pm$	0.85	&	11.85	$\pm$	0.15		\\
		&					&	3	&	25.21	$\pm$	0.38		&	1.77	$\pm$	0.64	&	11.992	$\pm$	0.095	\\
28677	&	 85 Tau         		&	1	&	14.65	$\pm$	0.88		&	2.0	$\pm$	1.2	&	11.77	$\pm$	0.22		\\
		&					&	2	&	18.59	$\pm$	0.41		&	1.16	$\pm$	0.72	&	11.85	$\pm$	0.18		\\
		&					&	3	&	23.43	$\pm$	0.83		&	2.0	$\pm$	1.2	&	11.82	$\pm$	0.33		\\
204188	&	IK Peg			&	1	&	$-$10.7	$\pm$	2.0		&	3.5	$\pm$	1.9	&	12.26	$\pm$	0.29		\\
		&					&	2	&	$-$6.97	$\pm$	0.26		&	3.26	$\pm$	0.33	&	13.10	$\pm$	0.13		\\
		&	WD0549+158		&	1	&	23.83	$\pm$	0.74		&	1.44	$\pm$	0.81	&	11.76	$\pm$	0.22		\\
		&	WD2004--605		&	1	&	$-$18.75	$\pm$	0.29		&	2.50	$\pm$	0.32	&	13.425	$\pm$	0.091	\\
		&					&	2	&	$-$12.3	$\pm$	1.6		&	1.92	$\pm$	0.92	&	12.27	$\pm$	0.22		\\
9672		&	 49 Cet         		&	1	&	11.0		$\pm$	1.6		&	4.74	$\pm$	0.65	&	12.69	$\pm$	0.21		\\
		&					&	2	&	13.65	$\pm$	0.15		&	2.50	$\pm$	0.28	&	13.27	$\pm$	0.10		\\
43940	&	HR2265			&	1	&	10.77	$\pm$	0.96		&	4.08	$\pm$	0.66	&	12.53	$\pm$	0.13		\\
		&					&	2	&	17.7		$\pm$	2.4		&	2.98	$\pm$	0.90	&	12.05	$\pm$	0.26		\\
		&					&	3	&	21.15	$\pm$	0.97		&	3.94	$\pm$	0.60	&	12.71	$\pm$	0.11		\\
137333	&	$\rho$ Oct			&	1	&	$-$9.27	$\pm$	0.15		&	2.58	$\pm$	0.18	&	12.772	$\pm$	0.015	\\
		&					&	2	&	$-$1.80	$\pm$	0.22		&	1.87	$\pm$	0.40	&	12.53	$\pm$	0.11		\\
		&					&	3	&	1.2		$\pm$	1.1		&	2.7	$\pm$	1.8	&	11.91	$\pm$	0.32		\\
		&					&	4	&	10.0		$\pm$	1.7		&	1.4	$\pm$	1.9	&	11.78	$\pm$	0.19		\\
3712	&	$\alpha$ Cas			&	1	&	$-$6.05	$\pm$	0.59		&	2.97	$\pm$	0.42	&	13.23	$\pm$	0.15		\\
		&					&	2	&	$-$1.8	$\pm$	1.6		&	2.1	$\pm$	1.5	&	11.83	$\pm$	0.60		\\
		&					&	3	&	8.95		$\pm$	0.38\tablenotemark{b}	&	3.47	$\pm$	0.49\tablenotemark{b}	&	12.435	$\pm$	0.057\tablenotemark{b}	\\
149382	&	HIP81145			&	1	&	$-$32.73	$\pm$	0.31		&	2.70	$\pm$	0.14	&	13.049	$\pm$	0.036	\\
		&					&	2	&	$-$25.28	$\pm$	0.31		&	3.89	$\pm$	0.38	&	13.657	$\pm$	0.097	\\
		&					&	3	&	$-$16.4	$\pm$	1.7		&	3.8	$\pm$	1.2	&	12.25	$\pm$	0.24		\\
		&	WD0621--376		&	1	&	8.76		$\pm$	0.35		&	2.26	$\pm$	0.76	&	12.385	$\pm$	0.069	\\
		&					&	2	&	15.93	$\pm$	0.33		&	2.81	$\pm$	0.48	&	12.733	$\pm$	0.068	\\
		&					&	3	&	22.41	$\pm$	0.30		&	2.02	$\pm$	0.37	&	12.274	$\pm$	0.059	\\
75747	&	RS Cha			&	1	&	$-$5.3	$\pm$	1.2		&	2.88	$\pm$	0.95	&	12.31	$\pm$	0.31		\\
		&					&	2	&	$-$1.84	$\pm$	0.41		&	2.40	$\pm$	0.35	&	12.89	$\pm$	0.14		\\
		&					&	3	&	10.22	$\pm$	0.39		&	3.39	$\pm$	0.15	&	13.400	$\pm$	0.054	\\
		&					&	4	&	16.8		$\pm$	1.2		&	2.9	$\pm$	1.0	&	12.54	$\pm$	0.18		\\
		&	IX Vel			&	1	&	3.63		$\pm$	0.23		&	3.33	$\pm$	0.28	&	12.866	$\pm$	0.071	\\
		&					&	2	&	4.06		$\pm$	0.30		&	1.20	$\pm$	0.43	&	12.49	$\pm$	0.17		\\
		&					&	3	&	16.1		$\pm$	1.3		&	3.3	$\pm$	1.1	&	12.96	$\pm$	0.22		\\
		&					&	4	&	18.26	$\pm$	0.89		&	3.04	$\pm$	0.32	&	13.09	$\pm$	0.16		\\
141569	&	HIP77542			&	1	&	$-$31.88	$\pm$	0.56		&	2.56	$\pm$	0.37	&	12.72	$\pm$	0.14		\\
		&					&	2	&	$-$27.55	$\pm$	0.15		&	2.18	$\pm$	0.93	&	13.444	$\pm$	0.046	\\
		&					&	3	&	$-$21.21	$\pm$	0.18		&	3.11	$\pm$	0.98	&	13.255	$\pm$	0.086	\\
		&					&	4	&	$-$12.89	$\pm$	0.23		&	2.87	$\pm$	0.53	&	13.68	$\pm$	0.11		\\
		&					&	5	&	$-$6.0	$\pm$	1.1		&	4.11	$\pm$	0.57	&	12.74	$\pm$	0.16		\\
		&					&	6	&			...				&	...				&	$<$11.7					\\
149730	&	R Ara			&	1	&	$-$24.15	$\pm$	0.63		&	3.37	$\pm$	0.58	&	12.88	$\pm$	0.21		\\
		&					&	2	&	$-$19.3	$\pm$	1.8		&	2.5	$\pm$	2.1	&	12.33	$\pm$	0.38		\\
		&					&	3	&	$-$14.28	$\pm$	0.57		&	2.5	$\pm$	1.6	&	12.48	$\pm$	0.25		\\
		&					&	4	&	$-$7.02	$\pm$	0.98		&	3.14	$\pm$	0.65	&	13.41	$\pm$	0.12		\\
		&					&	5	&	0.5		$\pm$	1.4		&	3.1	$\pm$	1.3	&	14.25	$\pm$	0.70		\\
		&					&	6	&	5.22		$\pm$	0.67		&	3.3	$\pm$	1.2	&	13.24	$\pm$	0.26		\\
\enddata
\tablenotetext{a}{heliocentric radial velocity}
\tablenotetext{b}{The third component of $\alpha$ Cas could not be modeled individually in the 2586 {\AA} line. Therefore, all final parameters for this component are the weighted means of the parameters in the 2600 {\AA} individual fit and the simultaneous fit.}
\end{deluxetable}

\clearpage


\begin{deluxetable}{llcccc}
\tablewidth{0pt}
\tabletypesize{\tiny}
\tablecaption{Fit Parameters for New \ion{Mn}{2} LISM Components\label{tabmn}}
\tablehead{& & Component & $v$\tablenotemark{a} & $b$ & $\log N_{\rm{MnII}}$ \\
HD No. & Other Name & Number & (km s$^{-1}$) & (km s$^{-1}$) & log(cm$^{-2})$\\}
\startdata
209100 & $\epsilon$ Ind &  & ... & ... &$<$11.8\\
115617 & 61 Vir & & ... & ... &$<$11.8 \\
114710 & $\beta$ Com&& ... & ... &$<$11.6 \\
       & WD1620--391&& ... & ... &$<$12.2 \\
72905  & $\pi^1$ UMa&& ... & ... &$<$12.0\\
217014 & 51 Peg&& ... & ... &$<$11.5 \\
120136 & $\tau$ Boo && ... & ... &$<$11.6\\
142373 & $\chi$ Her  && ... & ... &$<$11.5 \\
220140 & V368 Cep       && ... & ... &$<$12.5 \\
97334  & MN UMa         && ... & ... &$<$11.9\\
       & WD1337+705    && ... & ... &$<$12.3\\
222107 & $\lambda$ And  && ... & ... &$<$11.8 \\
180711 & $\delta$ Dra   && ... & ... &$<$11.8 \\
12230  & 47 Cas         && ... & ... &$<$11.9 \\
163588 & $\xi$ Dra      && ... & ... &$<$12.1 \\
216228 & $\iota$ Cep    && ... & ... &$<$12.0 \\
93497  & $\mu$ Vel      && ... & ... &$<$11.6 \\
149499 & V841 Ara       && ... & ... &$<$12.7 \\
210334 & AR Lac         && ... & ... &$<$12.0 \\
28911  & HIP21267       && ... & ... &$<$11.9 \\
28677  & 85 Tau         && ... & ... &$<$11.5 \\
204188 & IK Peg         && ... & ... &$<$11.5 \\
       & WD0549+158     && ... & ... &$<$12.2 \\
       & WD2004-605     && ... & ... &$<$12.2 \\
43940&HR2265&& ... & ... &$<$11.4\\
9672   & 49 Cet         &1& 10.47 $\pm$ 0.80 & 3.61 $\pm$ 0.66 & 11.345 $\pm$ 0.097 \\
&&2& 14.42 $\pm$ 0.82 & 3.41 $\pm$ 0.72 & 11.32 $\pm$ 0.16 \\
137333 & $\rho$ Oct     && ... & ... &$<$12.2 \\
3712   & $\alpha$ Cas   && ... & ... &$<$11.8 \\
149382 & HIP81145       &1& $-$31.28 $\pm$ 0.79 & 1.83 $\pm$ 0.82 & 11.27 $\pm$ 0.16 \\
&&2& $-$25.41 $\pm$ 0.91 & 3.9 $\pm$ 1.4 & 11.92 $\pm$ 0.19 \\
&&3& $-$14.3 $\pm$ 4.7 & 4.04 $\pm$ 0.84 & 11.64 $\pm$ 0.15 \\
       & WD0621-376     && ... & ... &$<$12.0 \\
75747  & RS Cha       && ... & ... &$<$11.5  \\
       & IX Vel         && ... & ... &$<$11.9\\
141569 & HIP77542       &1& $-$31.23 $\pm$ 0.69 & 1.98 $\pm$ 0.71 & 11.59 $\pm$ 0.13\\
&&2& $-$27.0 $\pm$ 1.4 & 1.68 $\pm$ 0.91 & 11.528 $\pm$ 0.081\\
&&3& $-$17.2 $\pm$ 1.6 & 3.09 $\pm$ 0.98 & 11.84 $\pm$ 0.35\\
&&4& $-$12.97 $\pm$ 0.48 & 2.93 $\pm$ 0.31 & 12.830 $\pm$ 0.090\\*
&&5,6& ... & ... &$<$11.5\\*
149730 & R Ara          &1,2,3,4& ... & ... &$<$11.8 \\
&&5& $-$1.4 $\pm$ 1.7 & 4.82 $\pm$ 0.73 & 12.70 $\pm$ 0.28 \\
&&6& 2.30 $\pm$ 0.38 & 3.48 $\pm$ 0.44 & 13.05 $\pm$ 0.19 \\
\enddata
\tablenotetext{a}{heliocentric radial velocity}
\end{deluxetable}


\setcounter{figure}{2}
\begin{figure}[t]
\centering
\centerline{\includegraphics{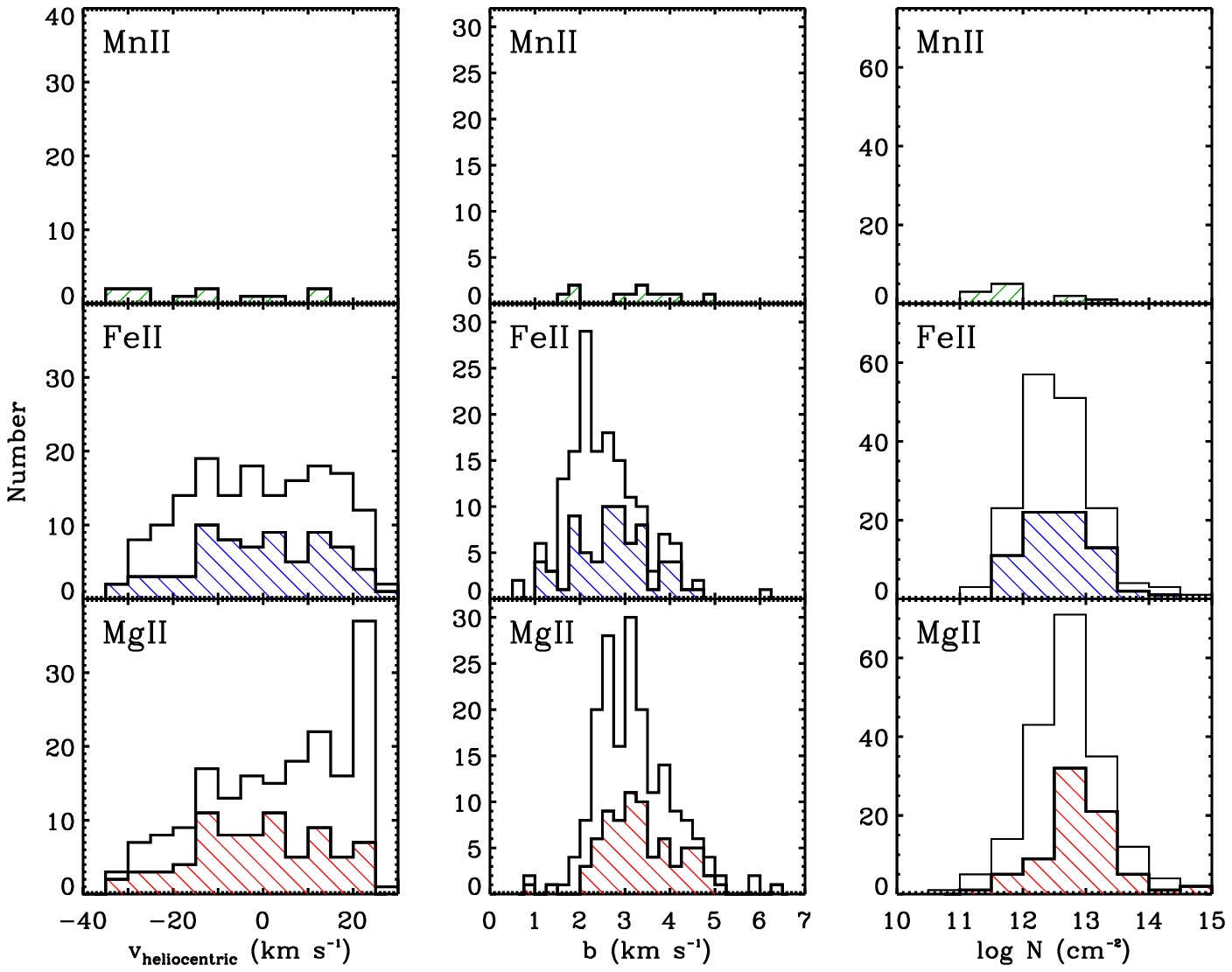}}
\caption{Distributions of heliocentric radial velocity ($V$), Doppler parameter ($b$), and column density ($N$) for the ions observed in this survey. The cross-hatched sample is the distribution of the SNAP survey, while the full LISM sample is shown by the clear distribution \citep{redfield02,redfield04sw}.  Bin sizes are 5 km s$^{-1}$, 0.25 km s$^{-1}$, and 0.5 log(cm$^{-2}$) respectively.  
\label{fighist}}
\end{figure}

\subsection{Velocity Distribution}

The radial velocity distribution (Figure~\ref{fighist}) contains the projected velocities of absorbing material in the heliocentric rest frame. Although the LISM clouds each have distinct motions, they move in the same general direction with similar velocities \citep{frisch02}. Therefore, most sight lines in this sample show clusters of components in velocity space rather than isolated components distributed at random velocities. Only in the longest sight lines do we detect a broader range of velocities. At these large distances, traversed clouds may include structures outside of the local interstellar environment, where the kinematics diverge from the general LISM flow vector.

\begin{figure}[!ht]
\centering
\centerline{\includegraphics{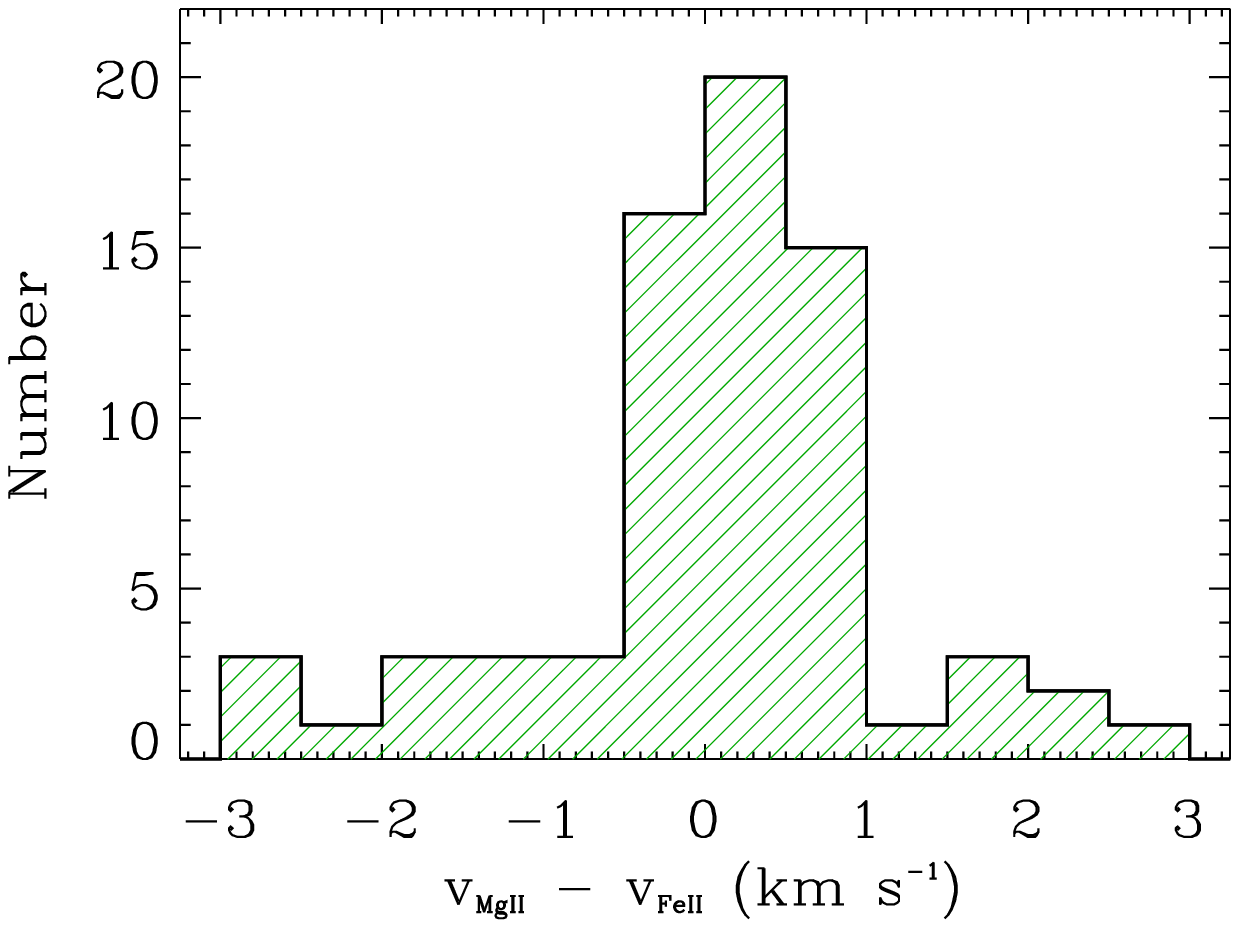}}
\caption{Distribution of differences between velocities of components measured with \ion{Mg}{2} and \ion{Fe}{2}. The distribution peaks around $\sim$0 km s$^{-1}$ and indicates that these ion pairs are well mixed and trace the same clouds. \label{vcompare}}
\end{figure}

The radial velocities in the distribution range from $-$32 to +25 km s$^{-1}$.  This span is consistent the bulk velocity of the warm LISM clouds (28.1 $\pm$ 4.6 km s$^{-1}$; \citealt{frisch11}), where an all-sky survey would sample the full range from positive values at this magnitude in the downwind direction to negative values at this magnitude in the upwind direction. When transformed to the reference frame of the Local Standard of Rest \citep{dehnen98}, the distribution is similar, although centered on 0 km~s$^{-1}$, slightly narrower ($\pm$20 km~s$^{-1}$), and more peaked.  The similarity of the general shape of the \ion{Mg}{2} and \ion{Fe}{2} distributions suggests that the ions are present in the same clouds. (Note that for the full LISM sample, the observed sight lines differ between \ion{Mg}{2} and \ion{Fe}{2}.  For example, \citet{redfield01} analyzed 18 sight lines toward the Hyades, which were observed in \ion{Mg}{2} and not in \ion{Fe}{2}, resulting in the noticeable spike in detections at $v_R \sim 21$ km~s$^{-1}$.)  If \ion{Mg}{2} and \ion{Fe}{2} are tracing the same material, then the $(v_{\rm{MgII}} - v_{\rm{FeII}})$ distribution should peak at 0 km s$^{-1}$, as indeed Figure \ref{vcompare} shows (mean$ = 0.06$ km~s$^{-1}$; standard deviation$ = 1.09$ km~s$^{-1}$).

\subsection{Doppler Parameter Distribution}

Mean values for the Doppler parameters and log column densities are listed in Table \ref{avg}. The well-known dependence of the Doppler parameter ($b$; [km~s$^{-1}$]) on temperature ($T$) and turbulent velocity ($\xi$) is:

\begin{equation}\label{eq:doppler} b^2= \frac{2kT}{m}+\xi^2=0.016629\frac{T}{A}+\xi^2,\end{equation}

\noindent where $A$ is the ion's atomic weight in atomic mass units, $k$ is the Boltzmann constant, and $m$ is the ion's mass. \ion{Mg}{2} experiences larger Doppler broadening on average because it is a lighter ion and more susceptible to the thermal contribution. Conversely, \ion{Mn}{2} and \ion{Fe}{2} should have roughly equivalent mean Doppler parameters because turbulence --- the dominant broadening mechanism for heavier ions --- is independent of atomic weight. The $\langle$$b$$\rangle$ value of \ion{Mn}{2} is 0.4 km s$^{-1}$ greater than that of \ion{Fe}{2}, but this discrepancy should be discounted due to the small number of \ion{Mn}{2} detections. \cite{redfield02} report $\langle$$b$(\ion{Fe}{2})$\rangle$ $\sim$ 2.4 km s$^{-1}$ with $\sigma_{\rm{FeII}}$ $\sim$ 1.0 km s$^{-1}$ and $\langle$$b$(\ion{Mg}{2})$\rangle$ $\sim$ 3.1 km s$^{-1}$ with $\sigma_{\rm{MgII}}$ $\sim$ 0.8 km s$^{-1}$. While both means are 0.3 km s$^{-1}$ lower than the values obtained here, the difference is not very significant.

\begin{deluxetable}{cccccc}
\tablewidth{0pt}
\tablecaption{Mean Values for Doppler Parameter and Log Column Density \label{avg}}
\tablehead{ & $\langle$$b$$\rangle$ & $\sigma_b$ & $\langle$log$N$$\rangle$ & $\sigma_{\rm{log}\emph{N}}$\\
Ion & (km s$^{-1}$) & (km s$^{-1}$) & log(cm$^{-2})$ & log(cm$^{-2})$}
\startdata
\ion{Mg}{2} & 3.36 & 0.90 & 12.89 & 0.58 \\
\ion{Fe}{2}  & 2.72 & 0.84 & 12.61 & 0.54 \\
\ion{Mn}{2} & 3.16 & 0.99 & 11.91 & 0.65 \\
\enddata
\end{deluxetable}

\begin{figure*}[!htb]
\centering
\centerline{\includegraphics[scale=0.7]{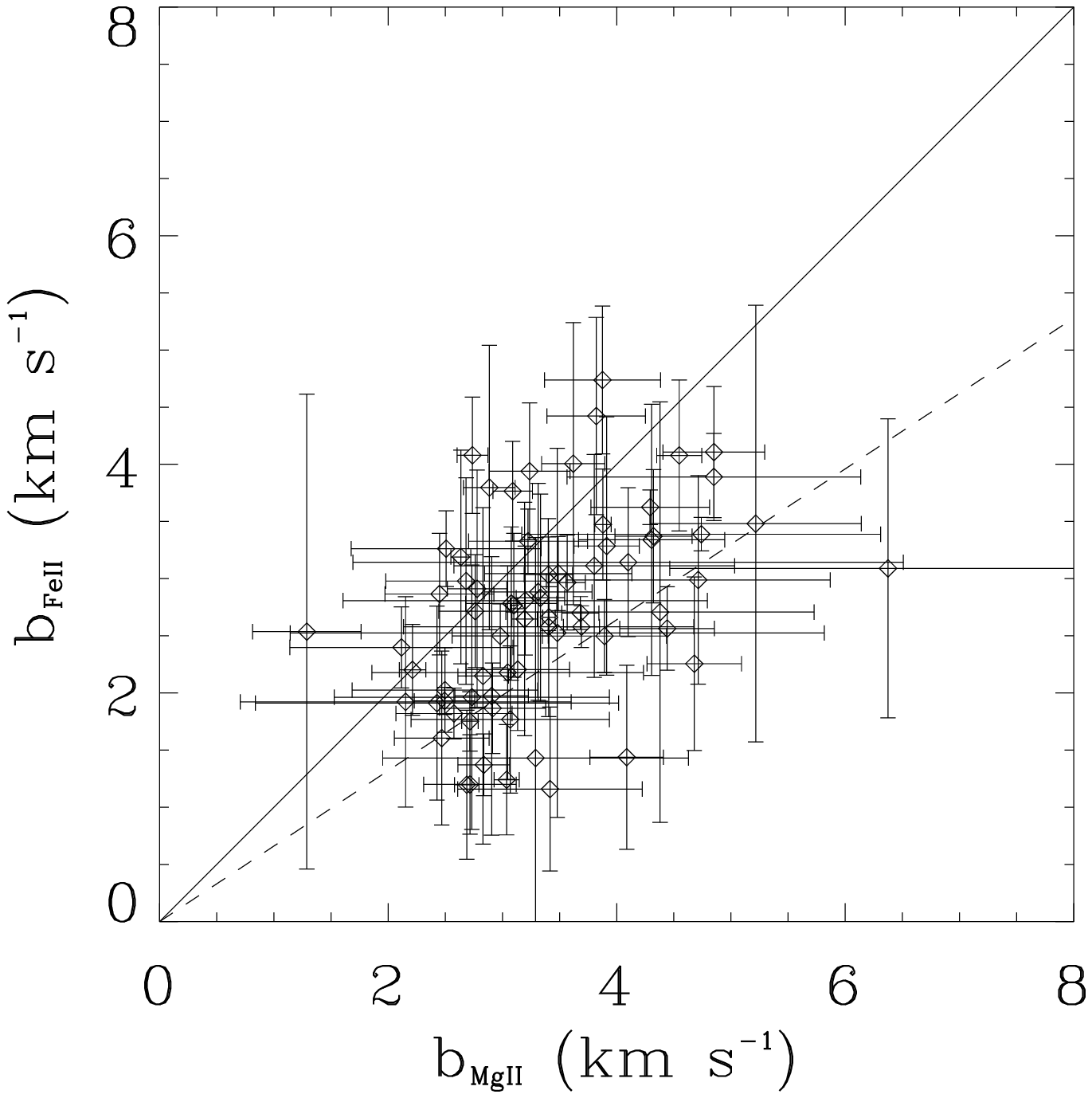}}
\caption{Comparison of \ion{Mg}{2} and \ion{Fe}{2} Doppler parameters of 71 individual components with 1$\sigma$ error bars. Solid line is the expected ratio for purely turbulent broadening; dashed line for purely thermal broadening. If a given \ion{Mg}{2} and \ion{Fe}{2} pairing truly belonged to the same cloud, the ratio should fall between the two limits.  Only a few points fall outside this zone, indicating that \ion{Mg}{2} and \ion{Fe}{2} are well mixed and trace the same material. \label{bvb}}
\end{figure*}

Figure \ref{bvb} compares Doppler widths of \ion{Mg}{2} and \ion{Fe}{2} absorptions for individual components. There are 71 pairings based on agreement in velocity. The solid line marks the ratio $b$$_{\rm{FeII}}$/$b$$_{\rm{MgII}}$ = 1, which would be the case if there were no thermal contribution to the line widths. If, on the other hand, the broadening were entirely thermal, a ratio of $b$$_{\rm{FeII}}$/$b$$_{\rm{MgII}}$ = 0.66 would be obtained.  When both broadening mechanisms contribute, the corresponding line width ratio should fall between the two lines. Indeed, 67 of the 71 components (94\%) fall within this zone allowing for the 1$\sigma$ error bars.  Three of the four remaining outliers have very weak \ion{Fe}{2} absorption that yield an artificially broad or narrow fit. Alternatively, these unphysical ratios might highlight components in \ion{Mg}{2} and \ion{Fe}{2} that do not originate from a common cloud, or might result from partial saturation of one of the species.


\subsection{Column Density Distribution}

The column density distribution is included in Figure \ref{fighist}, and the means and standard deviations for each ion are listed in Table \ref{avg}. \ion{Mg}{2} and \ion{Fe}{2} show similar column densities, but \ion{Mn}{2} is approximately an order of magnitude smaller. While the \ion{Mn}{2} sample is small, this difference still holds if one considers the average $\log N_{\rm{MnII}}$ upper limit of 11.8 $\log$(cm$^{-2})$. The similarity in \ion{Mg}{2} and \ion{Fe}{2} column densities is attributable to two factors: both elements have comparable cosmic abundances, and both ions are the dominant ionization stages in the LISM \citep{slavin08}. The lower \ion{Mn}{2} column densities are due in part to the much lower cosmic abundance of Mn (about two orders of magnitude lower than Mg and Fe). While Mn is not included in the ionization balance calculations of \citet{slavin08}, its first ionization potential is very similar to Mg and Fe, and so \ion{Mn}{2} is also likely the dominant ionization stage in the LISM.


\subsection{Number of Components Versus Distance}

The sight lines in this sample contain anywhere from one to six components. As would be expected, the number of components roughly correlates with the sight line length.  With observations of enough sight lines, the number of components, together with an accurate distance to the background star, can provide insight into the distribution of LISM clouds as a function of distance.  \citet{redfield04a} examined sight lines in 10 pc bins to judge how the average number of components per sight line changes with distance.  A uniform distribution of identically sized clouds should show a steady increase in the average cloud number per 10 pc increment. Instead, the distribution remains flat after 30 pc, suggesting that LISM clouds are concentrated close to the solar system. The \citet{redfield04a} sample, however, suffers from an under-sampling of sight lines approaching 100 pc. In addition, a comprehensive survey of distant clouds requires dense spatial sampling on the sky.


\begin{figure}[!htb]
\centering
\centerline{\includegraphics[scale=0.7]{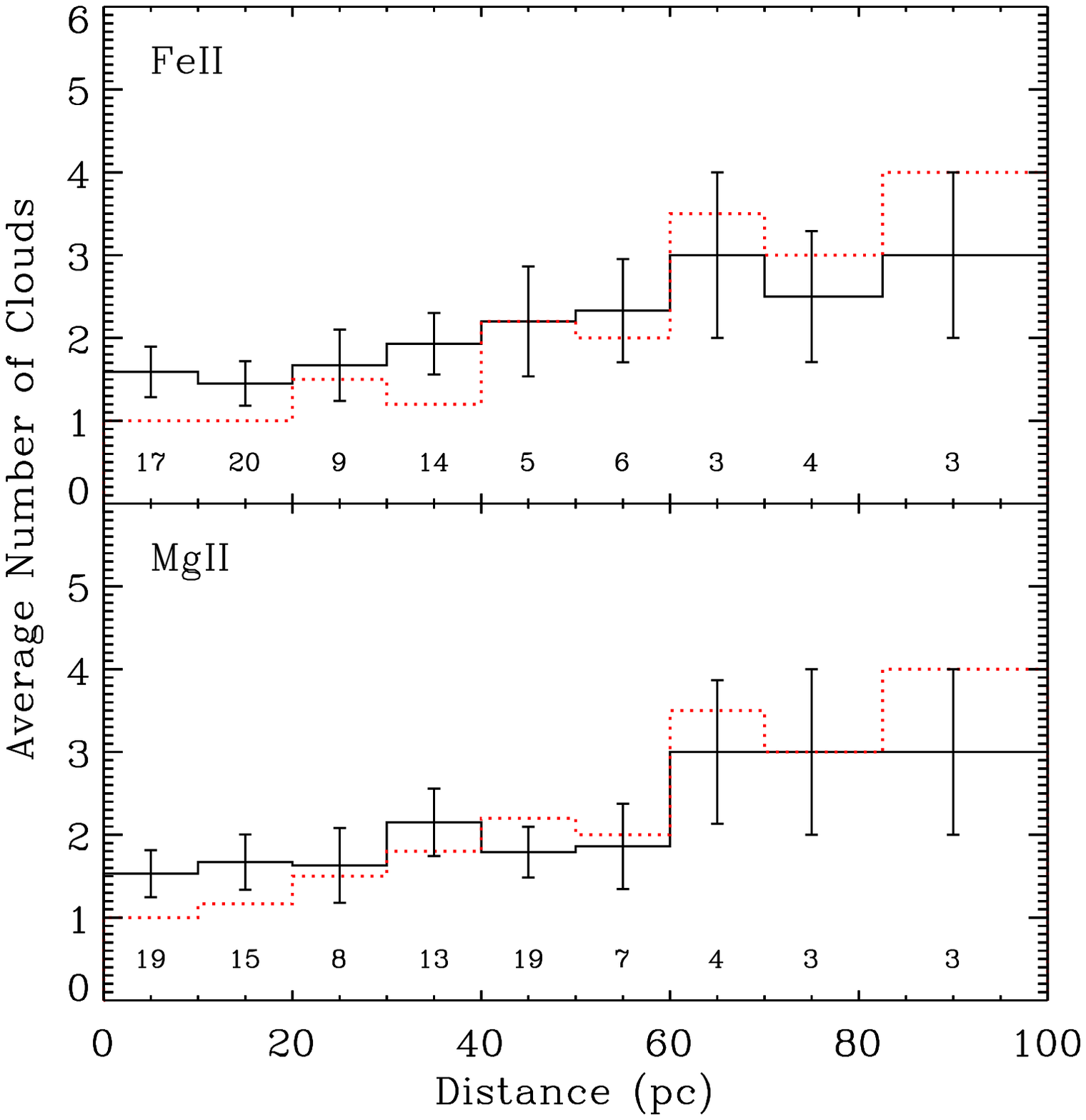}}
\caption{Distribution of the average number of detected absorption components per sight line in 10 pc bins. Errors are $\sqrt{N}$.  The solid black line shows averages of the \cite{redfield04a} data set combined with the SNAP sample. The red dotted line shows the distribution for the SNAP sample alone.  The numbers below each bin indicate the total number of sight lines in that bin. Both distributions suggest that the average cloud number remains fairly constant out to 50--70 pc, supporting the conclusion by \cite{redfield04a} that most clouds in the Local Bubble lie within 15 parsecs of the Sun. The rise beyond 70 pc is likely due to the onset of the Local Bubble edge in certain directions.  \label{colhist}}
\end{figure}

In an effort to improve this measurement, we combined the older sample with the new SNAP sight lines, which include five lines of sight longer than 70 pc as well as many more shorter ones. The new distribution (Figure \ref{colhist}) shows a slight positive trend within the first 50 pc, but the uncertainties are such that the rise probably is not significant. This consistency in the average number of absorbers indicates that most clouds located within 50 parsecs actually begin within 10 pc of the Sun. More measurements are needed to determine whether the slight rise is real. Between 50 and 70 pc, there is a jump in the average number of absorbers. This increase might be related to the onset of the closest edge the Local Bubble at $\sim$55--60 pc \citep{lallement03,welsh10}. The elevated level appears to continue out to 100 parsecs.


\section{Discussion}
\label{discussion.s}

The addition of 34 sight lines to the sample of heavy ions previously observed in the LISM opens up many avenues for further research, especially as these data are paired with the existing moderate or high-resolution FUV observations.  Fundamental properties, such as temperature, turbulence, ionization, abundances, and dust depletions of gas in the LISM can be measured by coupling together such  observations.  Serendipitous overlap with other areas of research often results in unexpected new directions for investigation. Included in this section are several examples of new results based on the new observations and a sampling of future research directions enabled by them.  We first use this sample to explore fundamental properties of the LISM including kinematics, small scale structure, and the thermal and turbulent properties of the gas.  Next we explore individual sight lines of particular interest, not only for their LISM absorption, but also for other sources of absorption, namely astrospheres and circumstellar disks.  Astrospheres have been detected around two of our stars.  We explore how our observations of the foreground ISM velocity structure impacts the interpretation of these detections.  Also, two stars in our sample have circumstellar disks.  We explore the possibility of absorption due to the disk and evaluate the ISM properties directly surrounding these stars.  
 
\subsection{Comparison with the LISM Kinematic Model}

One of the fundamental results of the kinematics of the LISM is that it can, to first order, be described by a single bulk flow vector.  This flow was first identified as coming from the direction of the Scorpio-Centaurus association by \citet{crutcher82} using high resolution ground-based \ion{Ti}{2} absorption spectra.  \citet{lallement86} recognized that while the interstellar flow is coming from a single general direction, multiple velocity vectors are needed to fully characterize the observed interstellar absorption toward nearby stars.  In a refinement of their analysis, \citet{lallement95} identified a nearby cloud, the G Cloud, in the Galactic center direction as distinct from the kinematics of the cloud directly surrounding the solar system, the Local Interstellar Cloud (LIC).  Using 96 sight lines, combining optical and UV observations, \citet{frisch02} evaluated the single bulk flow, but noted that significant deviations from this flow were apparent and identified seven LISM clouds.  As the number of high resolution LISM UV observations increased, spatially coherent discrepancies from the single bulk interstellar flow became evident \citep{redfield09}.  Within 15 pc of the Sun, \citet{redfield07lism4} identified 15 LISM clouds based on 157 sight lines, of which only 81 were optimal high-resolution NUV spectra. In order to solve for a single velocity vector of an LISM cloud, or test whether indeed a single bulk flow vector is sufficient to characterize the kinematics, one needs to correlate a large number of radial velocity measurements across an area of sky.   

Subsequent observations provide excellent tests to the predictions made by the kinematic model developed by \citet{redfield07lism4}.  For example, \citet{welsh10pasp} analyzed the absorption spectra of three B-stars at distances of $\sim$70 pc, finding a total of 11 individual absorption components.  They note the good agreement between the kinematic predictions based on the model by \citet{redfield07lism4}, and the observed velocities, where all five predicted clouds were detected in these three sight lines.  In addition, a cloud predicted to be near one of the sight lines, was also detected.  Of the remaining five detected components, the authors associated two with more distant interstellar structures (the Local Cavity, and the Loop I superbubble), and three components remained unidentified (a similar fraction remained unidentified in the \citet{redfield07lism4} analysis and may be associated with small, distant LISM clouds).  \citet{wood14} and \citet{woodpi1uma14} analyzed two new sight lines toward nearby stars and detected three absorption components in total.  In both cases, the \citet{redfield07lism4} kinematic model predicted absorption by the LIC, which was confirmed by the observations, and the third component remains unidentified.  The SNAP sample adds 34 new NUV sight lines within 100 pc, further increasing the total LISM sample size and enabling an additional and substantial test of the predictions made by the LISM kinematic model by \citet{redfield07lism4}.

Accompanying their paper, \cite{redfield07lism4} provided an online ``Kinematic Calculator"\footnote{http://lism.wesleyan.edu/LISMdynamics.html} that calculates the projections of all 15 cloud velocity vectors towards any direction in the sky. It also lists any clouds predicted to be along the line of sight or within $\sim$20$^{\circ}$, a rough estimate of the typical error in projected cloud boundaries given the modest sampling of the initial dataset. The first step in testing whether a velocity component is associated with any cloud is to compare that velocity to the radial velocities of all the clouds predicted along the line of sight. When the difference between the observed and predicted velocities is within 3$\sigma$ of zero, we consider the cloud to be a match. When more than one component along a line of sight meets that criterion, we select the one that agrees best.  The process continues until all the components have been associated with a known cloud, including those for which the sight line passes within 20$^{\circ}$ of the cloud boundary.  When a component's radial velocity matches that of a nearby cloud, it presents an opportunity to revise the cloud's boundaries. The boundaries were originally constructed by drawing contours around sight lines that showed spatial and kinematic similarities. The addition of more sight lines increases the ``resolution" of the cloud boundaries.  In the event that a component is incompatible with the velocities of all nearby clouds, it likely is the signature of a new, unidentified cloud. 

Table \ref{model} lists every velocity component detected in this sample and the cloud with which it best agrees. If the component does not match any clouds, it is labeled ``NEW." The label ``disk" denotes possible absorption by a circumstellar disk rather than the ISM (see Section~\ref{sec:disks}).  The listed velocities are the weighted means of the velocities measured in \ion{Mg}{2}, \ion{Fe}{2}, and \ion{Mn}{2}, unless one or both of the latter were not detected.  In some cases, the \ion{Mg}{2} or \ion{Fe}{2} velocity is not used in the calculation of the mean when the line is severely blended, saturated, or marginally detected.

\begin{deluxetable} {llccccc}
\tablewidth{0pt}
\tabletypesize{\tiny}
\tablecaption{Comparison with the LISM Kinematic Model \label{model}}
\tablehead{HD No.& Other &Distance & Component & $v$ & Cloud & $\sigma$\tablenotemark{a} \\
& Name & (pc) & Number & (km s$^{-1}$) & &}
\startdata
209100 & $\epsilon$ Ind & 3.62 & 1 & $-$10.83 $\pm$ 0.35 & LIC & 1.1 \\
115617 & 61 Vir & 8.56 & 1 & $-$14.74 $\pm$ 0.42 & NGP & 1.6 \\
114710 & $\beta$ Com&9.13&1& $-$6.00 $\pm$ 0.17 & NGP & 0.1 \\
& WD1620--391&13.2&1& $-$25.25 $\pm$ 0.28 & G & 1.2 \\
72905  & $\pi^1$ UMa&14.4&1&$ 13.29 \pm$ 0.24 & LIC & 1.0 \\
217014 & 51 Peg&15.6&1& $-$1.94 $\pm$ 0.31 & Eri & 0.7 \\
&&&2& 5.01 $\pm$ 0.55 & Hyades\tablenotemark{b} & 1.9 \\
120136 & $\tau$ Boo &15.6&1& $-$11.64 $\pm$ 0.21 & NGP & 2.0 \\
142373 & $\chi$ Her  &15.9&1& $-$12.70 $\pm$ 0.15 & NGP & 2.5 \\
220140 & V368 Cep       &19.2&1& 6.06 $\pm$ 0.22 & LIC & 0.4 \\
97334  & MN UMa         &21.9&1& 4.54 $\pm$ 0.19 & LIC & 0.5 \\
       & WD1337+705    &26.1&1& 1.67 $\pm$ 0.32 & LIC & 0.0 \\
222107 & $\lambda$ And  & 26.4&1& 0.24 $\pm$ 0.70 & NEW & \\
&&&2& 4.81 $\pm$ 0.24 & LIC & 1.2 \\
&&&3& 10.27 $\pm$ 0.21 & Hyades\tablenotemark{b} & 1.1 \\
180711 & $\delta$ Dra   & 29.9&1& $-$1.78 $\pm$ 0.14 & LIC & 0.2 \\
12230  & 47 Cas         & 33.2&1& 10.076 $\pm$ 0.071 & LIC & 0.2 \\
163588 & $\xi$ Dra      &34.5 &1& $-$13.72 $\pm$ 0.76 & Mic\tablenotemark{b} & 0.9 \\
&&&2& $-$6.27 $\pm$ 0.42 & LIC & 0.4 \\
216228 & $\iota$ Cep    & 35.3&1& 2.93 $\pm$ 0.18 & LIC & 1.6 \\
93497  & $\mu$ Vel      & 35.9&1& $-$5.61 $\pm$ 0.53 & G & 0.0 \\
&&&2& 0.2 $\pm$ 1.3 & Cet\tablenotemark{b} & 1.8 \\
149499 & V841 Ara       & 36.4&1&  $-$25.90 $\pm$ 0.63 & Aql\tablenotemark{b} & 1.7 \\
&&&2& $-$19.56 $\pm$ 0.91 & G & 1.2 \\
&&&3& $-$13.32 $\pm$ 0.68 & NEW \\
210334 & AR Lac         & 42.8 &1& $-$13.32 $\pm$ 0.34 & NEW & \\
&&&2& $-$0.63 $\pm$ 0.33 & LIC & 0.6 \\
28911  & HIP21267       & 44.7&1& 14.30 $\pm$ 0.26 & Hyades & 1.0 \\
&&&2& 20.3 $\pm$ 2.2 & Aur & 0.9 \\
&&&3& 23.83 $\pm$ 0.45 & LIC & 0.3 \\
28677  & 85 Tau         & 45.2&1& 13.81 $\pm$ 0.77 & Hyades & 0.4 \\
&&&2& 18.6 $\pm$ 1.1 & Aur & 2.3 \\
&&&3& 23.26 $\pm$ 0.36 & LIC & 0.3 \\
204188 & IK Peg         &46.4&1& $-$11.6 $\pm$ 1.3 & Eri & 1.1 \\
&&&2& $-$6.72 $\pm$ 0.24 & LIC\tablenotemark{b} & 0.6 \\
       & WD0549+158     & 49&1& 22.58 $\pm$ 0.81 & LIC & 0.8 \\
       & WD2004-605     & 58&1& $-$18.73 $\pm$ 0.28 & Vel & 2.3 \\
&&&2& $-$12.06  $\pm$ 0.79 & LIC\tablenotemark{b} & 1.2 \\
9672   & 49 Cet         &59.4&1& 10.15 $\pm$ 0.70 & LIC & 0.6 \\
&&&2& 13.69 $\pm$ 1.4 & disk & \\
43940&HR2265   & 61.9&1&  11.13 $\pm$ 0.23 & Blue & 0.1 \\*
&&&2& 18.21 $\pm$ 0.69 & Dor\tablenotemark{b} & 1.5 \\*
&&&3& 22.28 $\pm$ 0.64 & NEW & \\*
137333 & $\rho$ Oct     & 66.1&1& $-$8.82 $\pm$ 0.59 & G & 0.6 \\*
&&&2& $-$1.9 $\pm$ 1.0 & Blue\tablenotemark{b} & 1.0 \\*
&&&3& 3.2 $\pm$ 3.4 & Aql\tablenotemark{b} & 0.7 \\*
&&&4& 9.61 $\pm$ 0.57 & NEW & \\*
3712   & $\alpha$ Cas   & 70.0&1& $-$6.49 $\pm$ 0.36 & NEW & \\*
&&&2& $-$2.4 $\pm$ 0.44 & NEW & \\*
&&&3& 9.24 $\pm$ 0.30 & LIC & 0.7 \\*
149382 & HIP81145       & 73.9&1& $-$32.73 $\pm$ 0.31 & G & 3.1 \\*
&&&2& $-$25.28 $\pm$ 0.31 & Mic & 0.1 \\*
&&&3& $-$16.4 $\pm$ 1.7 & Leo\tablenotemark{b} & 2.5 \\*
& WD0621-376     & 78&1&  9.11  $\pm$ 0.31 & Blue & 1.0 \\*
&&&2& 16.00  $\pm$ 0.28 & Dor\tablenotemark{b} & 0.6 \\
&&&3& 22.41  $\pm$ 0.29 & NEW & \\
75747  & RS Cha     &92.9 &1& $-$4.86 $\pm$ 0.12 & G & 0.6 \\
&&& 2& $-$1.40 $\pm$ 0.11 & Vel & 1.7 \\
&&&3& 10.64 $\pm$ 0.12 & NEW & \\
&&&4& 17.27 $\pm$ 0.11 & NEW & \\
& IX Vel         &96.7 &1& 2.268 $\pm$ 0.062 & LIC\tablenotemark{b} & 2.4 \\
&&&2& 4.91 $\pm$ 0.37 & G\tablenotemark{b} & 1.1 \\
&&&3& 16.44 $\pm$ 0.32 & Vel\tablenotemark{b} & 0.7 \\
&&&4& 20.80 $\pm$ 0.73 & Cet\tablenotemark{b} & 2.2 \\
141569 & HIP77542       &116 &1& $-$31.88 $\pm$ 0.56 & NEW & \\
&&&2&                                          $-$27.55 $\pm$ 0.15 & G & 0.7 \\
&&&3& $-$21.21 $\pm$ 0.18 & Leo\tablenotemark{c} & 0.4 \\
&&&4& $-$12.89 $\pm$ 0.23 & NEW & \\
&&&5& $-$6.0 $\pm$ 1.1 & disk & \\
&&&6& 2.84 $\pm$ 0.44 & NEW & \\
149730 & R Ara          &124 &1&  $-$24.15 $\pm$ 0.63 & Aql\tablenotemark{b} & 0.0 \\
&&&2& $-$19.3 $\pm$  1.8 & G & 1.0 \\
&&&3& $-$14.28 $\pm$ 0.57 & NEW & \\
&&&4& $-$7.02 $\pm$ 0.98 & Blue\tablenotemark{b} & 0.2 \\
&&&5& $-$2.5 $\pm$ 1.8 & NEW & \\
&&&6& 2.13 $\pm$ 0.47 & NEW & \\
\enddata
\tablenotetext{a}{$\sigma = | v_{\rm obs} - v_{\rm pred} | / \sqrt{\sigma^2_{\rm obs} + \sigma^2_{\rm pred}}$, where the predicted cloud velocities and errors are taken from the velocity vectors calculated by \citet{redfield07lism4}.}
\tablenotetext{b}{This sight line is slightly outside the nominal cloud boundary determined in \citet{redfield07lism4}.}
\tablenotetext{c}{The Oph Cloud is an alternative cloud assignment for this sight line, with $\sigma = 0.0$.}
\end{deluxetable}

\clearpage

For the new sample of 34 sight lines, the kinematic model developed by \citet{redfield07lism4} predicts 40 absorption components spatially and dynamically consistent with the 15 known LISM clouds.  These absorbers have measured velocities within 3$\sigma$ of the predicted cloud radial velocity 100\% of the time (i.e., 40/40).  The excellent agreement provides strong support for the existence of spatially distinct, kinematic groups that can be characterized by simple bulk flow velocity vectors.  The SNAP sample is approximately randomly distributed, so the two nearest clouds that subtend the largest area on the sky (LIC and G) have the largest numbers of new detections, 19 and 9, respectively.  These are substantial increases in the total number of sight lines detected through these clouds.  \citet{redfield07lism4} used 79 sight lines for the LIC (so the new data represent a 24\% increase), and 21 sight lines for the G Cloud (a 43\% increase).  In addition to the 40 velocity components consistent with the boundaries of the 15 clouds, there are also 18 additional velocity components that were successfully predicted by the kinematic model to be within $\lesssim$20$^{\circ}$ of a cloud boundary.  These 18 velocity components will lead to refinements and improved spatial resolution in the LISM cloud boundaries.

Further, there were 18 components that are unaffiliated with known clouds.  Two are candidates for circumstellar absorption (see Section~\ref{sec:disks}), leaving 16 new unaffiliated radial velocities, half of which are detected in the lines of sight toward stars at distances $>$80 pc.  This subsample represents $\sim$21\% of the total number of detected absorbers, which is similar to the 19\% of the total absorbers that are unaffiliated in the original kinematic model by \citet{redfield07lism4}.  These absorbers likely are associated with more distant LISM clouds that subtend smaller angles, and therefore have fewer measurements with which to determine a velocity vector.  However, as the sample grows, it will become possible to characterize the dynamics of even these more distant LISM clouds.

\subsection{Small-scale Structure \label{secsss}}

The LISM is an inhomogeneous structure \citep{frisch91,diamond95,redfield07lism4}. Depending on the direction of observation, total column densities can vary by more than an order of magnitude over the same distance \citep{redfield02}. These changes are apparent over large angular scales, but to examine the finer structure requires sight lines with small angular separation.  Variations in the general ISM on sub-parsec scales have been found in previous studies of wide binaries \citep{meyer96,watson96} and high proper motion pulsars \citep{frail94,stanimirovic10}.  However, given the small number of observations, there are not many opportunities to make measurements of small scale structure in the LISM.  In their study of \ion{Mg}{2} absorption lines in the spectra of 18 stars in the Hyades with angular separations from 0.6--33$^{\circ}$, \citet{redfield01} found a smooth variation in column density, consistent with a large-scale homogeneous morphology of the LIC.  In the present survey, three pairs of stars provide spatial information on a scale of $<$3$^{\circ}$.

\begin{figure}[!ht]
\centering
\includegraphics[scale=0.69]{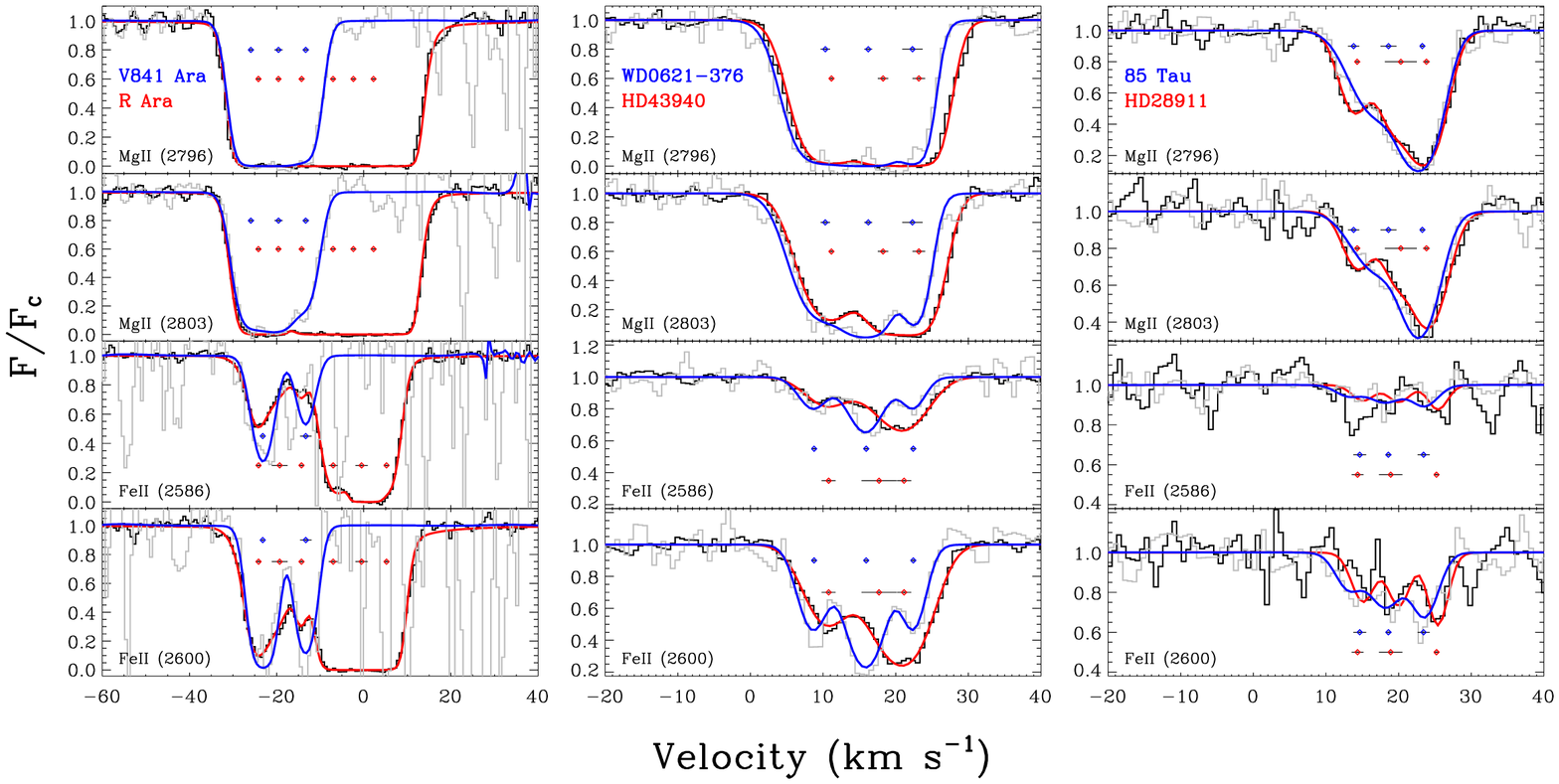}
\caption{Comparison of interstellar absorption in three pairs of sight lines with angular separation $<$ 3$^{\circ}$. The observed spectra for each star are shown in histogram mode, one in black and the second in grey.  At left are normalized spectra of R Ara (124 pc) and V841 Ara (36.4 pc), which have an angular separation of 0.5$^{\circ}$. At center are HD43940 (61.9 pc) and WD0621-376 (78 pc), with angular separation of 1.3$^{\circ}$. At right are 85 Tau (45.2 pc) and HD28911 (44.7 pc), separated by 2.6$^{\circ}$. Points depict velocity centroids (with 1$\sigma$ error bars) for each component fit to the respective line. \label{sss}}
\end{figure}

V841 Ara and R Ara are separated by only 0.5$^{\circ}$, but their distances from the Sun are 37.1 pc and 124 pc respectively. The R Ara sight line shows three extra components, suggesting that three clouds are located between 37.1--124 pc, including absorption from the Local Bubble boundary (Figure \ref{sss}). It also is possible that the cloud boundary is located in the 0.5$^{\circ}$ between the sight lines, although the physical separation would then correspond to only 0.32 pc at the distance of V841 Ara (and only 0.03 pc or 6300 AU at a distance of 3.5 pc, the approximate distance of the Aql Cloud, which matches component 1 for both lines of sight). The pair shows nearly identical velocity structure for the three bluest components, two of which have been identified as the Aql Cloud and the G Cloud. The $\sim$$-$14 km s$^{-1}$ component seen in both sight lines appears to be a new cloud located within 37.1 pc.  The other fit parameters (i.e., $b$-value and column density) also match the Aql and G clouds quite well.  For the Doppler parameter, all components for both \ion{Mg}{2} and \ion{Fe}{2} agree to within 3$\sigma$ of the Aql and G cloud velocities, with an average discrepancy of only 0.7$\sigma$.  Likewise for column density, all agree to within 3$\sigma$ with an average discrepancy of only 1.5$\sigma$.  Two of the three components seen only in R Ara likely belong to unidentified clouds located beyond 37.1 pc.  The third component is identified with the Blue Cloud which appears to traverse the sight line of R Ara, but not that of V841 Ara.

With a separation of 1.3$^{\circ}$, HD43940 (61.9 pc) and WD0621-376 (78 pc) show three similar absorption components that agree in velocity (Figure \ref{sss}). The two bluest components match the model predictions for the Blue and Dor Clouds. The third component is likely an unidentified cloud seen in both sight lines. The physical separation of clouds along the lines of sight is 1.4 pc at the distance of HD43940, 0.3 pc at the approximate distance of the Dor Cloud (11.7 pc), and only 0.06 pc or 12000 AU at the approximate distance of the Blue Cloud (2.6 pc).  The other fit parameters are quite similar to the Dor and Blue clouds, except for the $\sim$22 km~s$^{-1}$ component in \ion{Fe}{2}.  This component is discrepant in column density by 3.5$\sigma$ and in Doppler parameter by 2.7$\sigma$.  The disagreement can be easily seen in Figure~\ref{sss}, which shows that the HD43940 sight line has a markedly broader and deeper absorption.  This component was previously unidentified, which makes it likely that it is a distant cloud, and therefore possible that a significant variation is seen across a small angular distance.  The \ion{Mg}{2} absorption for this component does not show the same variation, but the absorption is near saturation making it difficult to identify subtle variations.  The other components all agree to within 3$\sigma$, with an average variation in Doppler parameter of only 0.7$\sigma$ and in column density of only 1.2$\sigma$.  Since no new component appears in the longer sight line, no new clouds with detectable column densities are located in the 16.1 pc span between the stars.

The pair of stars, HD28911 (44.7 pc) and 85 Tau (45.2 pc), separated by 2.6$^{\circ}$, are remarkable in that they are so close both in terms of their angular separation and distance.  These two sight lines are the only ones for which the kinematic model of \citet{redfield07lism4} predicts three different clouds along a line of sight.  Both sight lines show absorption at the expected radial velocity for all three clouds, with no additional components detected, implying that there are no LISM clouds beyond $\sim$5 pc (the approximate distance of the Hyades Cloud and the most distant of the three clouds), along these lines of sight out to 44.7 pc.  The fit parameters for all three components match the LIC, Hyades, and Aur clouds very well.  The Doppler parameters also fit those three clouds with discrepancies within 0.7$\sigma$, and the average variation agrees within 0.4$\sigma$.  For column density, all measurements agree to within 1.8$\sigma$, and the average variation agrees within 0.6$\sigma$.  While the physical separation of the pair at the distance of the closer star, HD28911, is only 2.0 pc, all three components are identified with LISM clouds.  At the approximate distance of the Hyades Cloud (5.0 pc), which matches component 1 for both sight lines, the physical separation of the lines of sight is 0.23 pc and for component 2, which matches with the Aur Cloud (3.5 pc), the physical separation is 0.16 pc.  Component 3 is associated with the LIC, which directly surrounds the solar system.  Based on the \citet{redfield00} morphological model of the LIC, and assuming an average \ion{H}{1} density of 0.2 cm$^{-3}$, the distance to the edge of the LIC is only 2.0 pc.  Therefore, the maximum physical separation of the sight lines at this distance is only 0.09 pc or 19000 AU. 

Based on these three close pairs, we find little evidence for significant small scale structure in the dominant ions of the warm partially ionized clouds of the LISM.  The predicted kinematic variations along such small separations is negligible.  We find that sight lines are identical in measured radial velocity, Doppler width, and column density down to scales of order 10,000 AU.

\subsection{Temperature and Turbulence}

Combining absorption measurements of multiple ions makes it possible to separate the thermal and turbulent contributions to the observed line widths.  With increasing ion mass, the contribution of thermal broadening to the Doppler parameter drops and the relative contribution of turbulence and unresolved cloud components increases. High spectral resolution observations are essential for the heavy ions (e.g., \ion{Fe}{2} and \ion{Mg}{2}).  On the other hand, medium resolution spectra are often suitable for the light ions (e.g., \ion{H}{1}, \ion{D}{1}, and \ion{C}{2}) located in the FUV.  The new sample of \ion{Mg}{2} and \ion{Fe}{2} Doppler widths can be combined with archival FUV data along the same sight lines to separate the temperature and turbulent velocity contributions. With a similar data set, \cite{redfield04b} measured the temperature ($T$) and turbulent velocity ($\xi$) for 50 individual absorbers.  Their measurements yielded a weighted mean LISM gas temperature ($T$) of 6680 K ($\sigma$ = 1490 K) and weighted mean turbulent velocity ($\xi$) of 2.24 km s$^{-1}$ ($\sigma$ = 1.03 km s$^{-1}$). 

\vspace{10mm}
\begin{figure}[!ht]
\centering
\includegraphics[scale=0.6, angle=90]{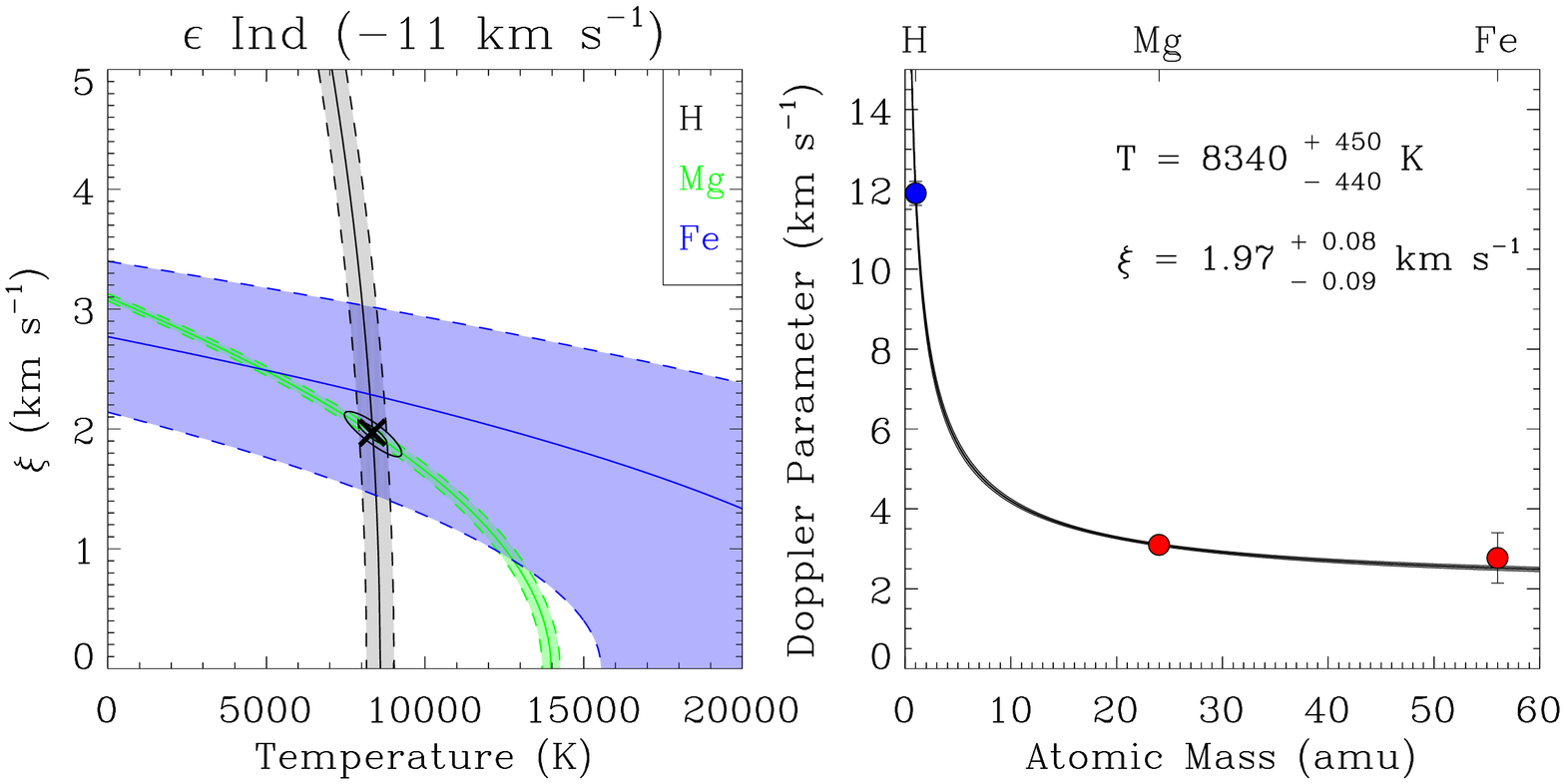}
\caption{Determination of temperature and turbulent velocity for LISM absorption observed toward $\epsilon$ Ind. The decomposition requires a light ion such as \ion{H}{1} and a minimum of one other ion at least as heavy as \ion{Mg}{2}. In the left panel, the best-fit Doppler parameter for each ion is the solid curve (dashed lines are $\pm$1$\sigma$), color coded according to the legend at the right. A black cross marks the best fit $T$ and $\xi$ given the $b$ values of all the ions involved. Surrounding the $\times$ are $\pm$1$\sigma$ and $\pm$2$\sigma$ error contours. In the right panel, the Doppler parameter is plotted against atomic mass, with corresponding species labeled at the top edge of the plot. The best-fit solution is shown (with $\pm$1$\sigma$ errors) in gray. \label{eindfig}}
\end{figure}

Here, we present a new calculation of $T$ and $\xi$ for the LISM absorption along the line of sight toward $\epsilon$ Ind (3.62 pc), see Figure \ref{eindfig}.  The single absorption component is kinematically associated with the LIC.  The \ion{H}{1} line width is taken from \citet{wood96}, who assumed a single component LISM absorption profile.  Our high-resolution NUV observations support that assumption and now enable the first measurement of $T$ and $\xi$ for the LISM along that sight line.  We find $T = 8340^{+450}_{-440}$ K and $\xi = 1.97^{+0.08}_{-0.09}$ km~s$^{-1}$, which agree well with the mean values for the LIC ($T$ = 7500 $\pm$ 1300 K, $\xi$ = 1.62 $\pm$ 0.75 km s$^{-1}$) obtained by \citet{redfield07lism4} using the 19 sight lines originally discussed in \citet{redfield04b}.  This result is dominated by \ion{Mg}{2} with its high precision line width measurement, and the (broader) \ion{H}{1} width.  Together with measurements of \ion{D}{1} or \ion{H}{1} absorption by \citet{wood00,wood05sup}, our new sample makes it possible to derive temperatures and turbulent velocities for V368 Cep, 61 Vir, $\chi$ Her, and $\mu$ Vel.  In the case of $\mu$ Vel, the previous assumption of only a single LISM absorber used in the analysis of the medium-resolution data turns out not to be valid as this sight line requires at least two LISM components.  All told, the present sample --- with 76 total absorbers and archived FUV observations --- has the potential to more than double the number of temperature and turbulent velocity measurements for the LISM and to continue developing a more refined inventory of the local interstellar environment.

\subsection{Astrospheres}

Knowledge of the LISM environment around a star is essential for a complete understanding of its astrosphere.  An astrosphere is analogous to the Sun's heliosphere, and marks the interface between the outward flow of the stellar wind and the inward pressure of the surrounding ISM. It can expand or contract depending on the density of the ISM as well as the strength of the stellar wind. \cite{wyman13} observed the LISM in the direction of the Sun's historical line of motion in order to investigate how it might have impacted the heliosphere and the modulation of the Galactic cosmic ray flux at 1 AU. The ISM environment through which a star is passing might achieve densities great enough to compress the astrosphere to within the orbit of planets, exposing their atmospheres directly to the ISM and the full brunt of the Galactic cosmic ray flux.

An astrosphere is detected by the spectral signature of its ``hydrogen wall." When cold ISM neutrals interact with protons from the hot solar wind, they charge exchange, producing an abundance of decelerated neutral hydrogen atoms \citep{wood04b}. This heated gas builds up at the interface of the stellar wind and the ISM, producing a deep, broad Ly$\alpha$ absorption feature. The astrospheric feature is often highly saturated and difficult to differentiate from interstellar H absorption blue-shifted relative to the ISM absorption. Observing heavy ions (e.g., \ion{Fe}{2} and \ion{Mg}{2}) in the same direction provides important constraints on the analysis of heliospheric and astrospheric \ion{H}{1} absorption by measuring the number LISM components and their radial velocities.

Observations of Ly$\alpha$ towards $\epsilon$ Ind and $\lambda$ And led \cite{wood96} to conclude that, for both sight lines, an astrospheric \ion{H}{1} absorption component was necessary to explain the width and velocity discrepancies between the \ion{H}{1} and \ion{D}{1} absorption lines. For each sight line, they identified one LISM component and an astrospheric component. Assuming $\xi$ = 1.2 km s$^{-1}$, they measured a LISM temperature $T$ = 8500 $\pm$ 500 K for $\epsilon$ Ind and $T$ = 11,500 $\pm$ 500 K for $\lambda$ And. The $\lambda$ And LISM temperature is high, suggesting that the broad Ly$\alpha$ absorption feature might contain blends.


\begin{figure}[!ht]
\centering
\includegraphics[scale=0.5, angle=90]{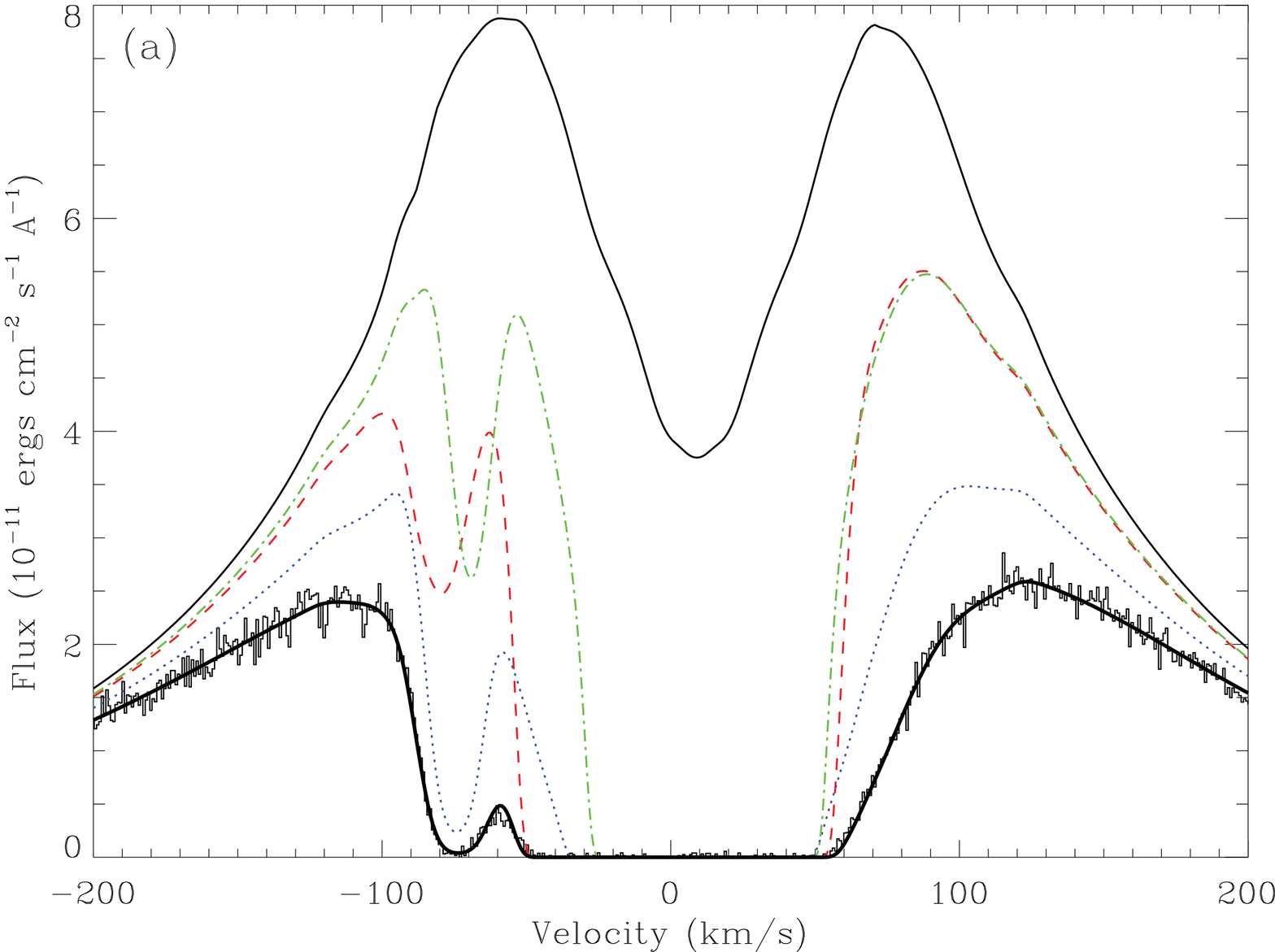}
\includegraphics[scale=0.5, angle=90]{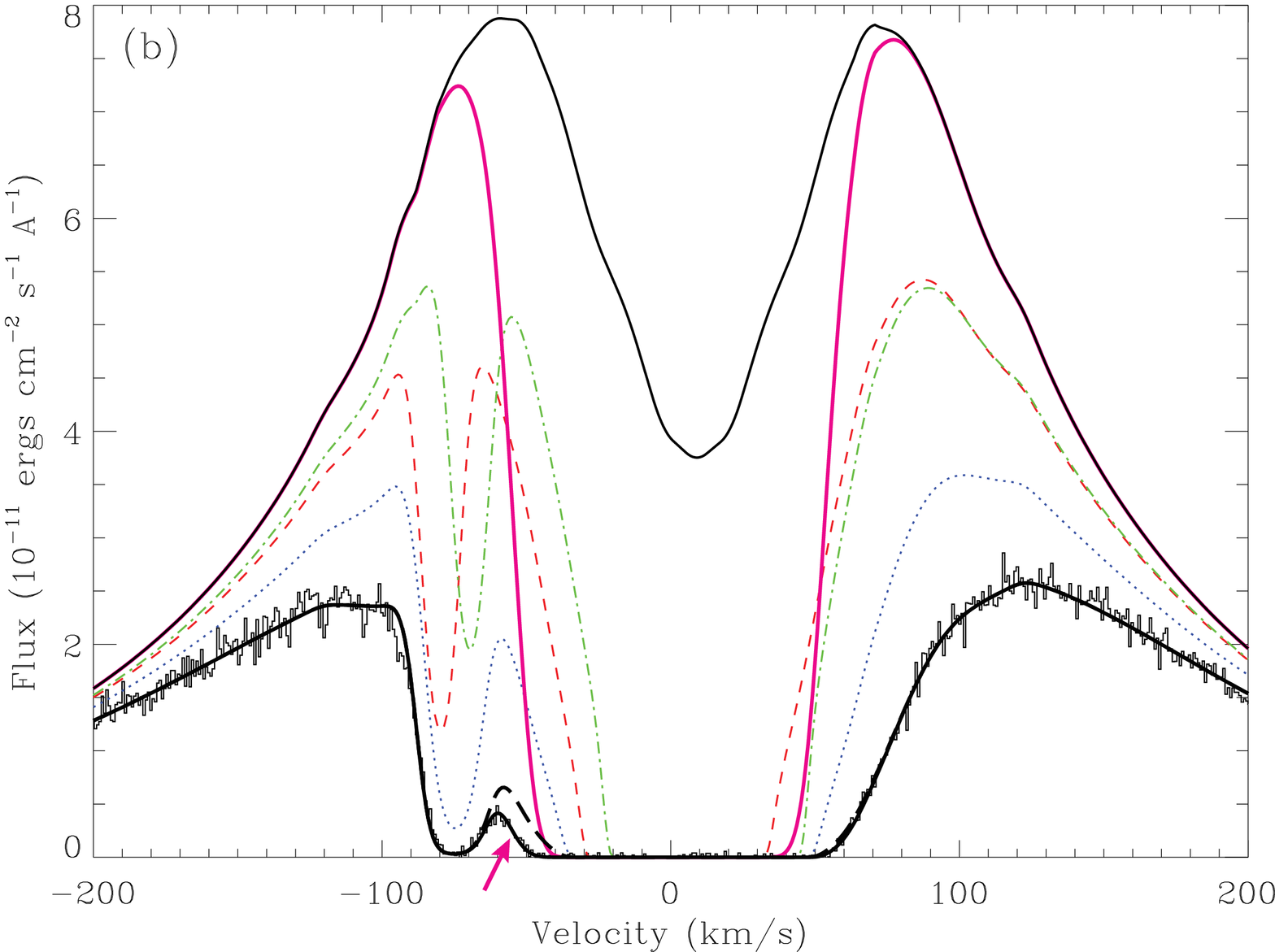}
\caption{\label{fit13} (a) Modeling of the $\lambda$ And Ly$\alpha$ line incorporating three ISM components (red, blue, and green curves for components 1--3, respectively) without an astrospheric component.  The final fit has a reduced $\chi^2_\nu$ = 1.21. The ISM-only model has some trouble matching the slope along the blue side of the saturated \ion{H}{1} absorption feature.  (b) Fit of the same Ly$\alpha$ line incorporating three ISM components (red, blue, and green curves for components 1--3, respectively) but with an astrospheric component (magenta line and the impact on the observed Ly$\alpha$ profile indicated by the magenta arrow).  Note that the blue side of the saturated \ion{H}{1} absorption is more convincingly matched with the addition of the astrospheric component (solid black line) compared to no astrospheric component (dashed black line).  The final fit is slightly improved with a reduced $\chi^2_\nu$ = 1.19. }
\end{figure}



The original $\epsilon$~Ind and $\lambda$~And analyses were performed
without knowledge of the ISM velocity structure.  However, we have now observed \ion{Mg}{2}
and \ion{Fe}{2} absorption for these lines of sight and determined the velocity structure.  This provides
motivation to revisit the Ly$\alpha$ analyses.  As
for $\epsilon$~Ind, the sight line contains only one absorption
component for both \ion{Mg}{2} and \ion{Fe}{2}, meaning that the analysis
of \citet{wood96} assuming just one ISM component requires no
revision.  The Doppler parameters indicate 
$\xi=1.97^{+0.08}_{-0.09}$ km~s$^{-1}$, a higher value than they
assumed, although the temperature measurement remains consistent.  The sight line toward 61 Vir, with a detection of both heliospheric and
astrospheric absorption is another example where the assumption of a
single LISM absorber was confirmed by the new high-resolution NUV
observations.  The heliospheric and astrospheric analysis performed by
\citet{wood05sup} on that star is therefore strengthened.

However, our examination of the \ion{Mg}{2} and \ion{Fe}{2} absorption lines towards
$\lambda$~And reveals three LISM components,
not one.  Therefore, the one-component Doppler parameter determined solely through
Ly$\alpha$ absorption is artificially enhanced, leading to an
overestimation of the LISM $T$ and $\xi$.  Furthermore, it is possible
that consideration of the complex velocity structure could obviate
the need for an astrospheric absorption component in the Ly$\alpha$
line.  In light of these new results, a reconsideration of the Ly$\alpha$
analysis of $\lambda$~And was warranted.

Figure~\ref{fit13} shows the Ly$\alpha$ spectrum of $\lambda$~And.  The
extremely broad \ion{H}{1} absorption is centered at about 10 km~s$^{-1}$,
with narrower deuterium (\ion{D}{1}) absorption at $-75$ km~s$^{-1}$.  The
methodology for Ly$\alpha$ modeling has been described in detail by \citet{wood05sup}.  
In short, the \ion{D}{1} and \ion{H}{1} absorption are fitted
simultaneously, with the \ion{D}{1} and \ion{H}{1} velocities and Doppler parameters
forced to be self-consistent.  In addition, we assume a column
density ratio of D/H$ = 1.56\times 10^{-5}$ for all components (the generally accepted ratio within the Local Bubble, but note that D/H
was not fixed in this manner in the original analysis of \citealt{wood96}).  
We assume that the velocity spacing of the three ISM
components of \ion{H}{1} is the same as that of the \ion{Mg}{2} components.  The background stellar Ly$\alpha$ profile is similar to that of \citet{wood96}, 
but with some minor modifications to achieve better fits given the fixed D/H ratio noted above.

We modeled the Ly$\alpha$ line profile with and without an astrospheric contribution.
Figure~\ref{fit13}a illustrates the best ISM-only fit, which assumes the same column
density ratios for the three components as derived from \ion{Mg}{2} but varies the
Doppler parameters.  The reduced $\chi^2_{\nu}=1.21$, but there is
a noticeable systematic discrepancy with the observed profile in the
$[-60,-50]$ km~s$^{-1}$ velocity range in Figure~\ref{fit13}a, and the Doppler
parameter of the most blueshifted ISM velocity component (Component~1)
required a very large ISM cloud temperature ($T\sim 18,000$~K).
A central issue regarding the acceptability of this fit is the
plausibility of such a high temperature for Component~1.  This
component seems to be detected in \ion{Ca}{2} for the very nearby target,
$\kappa$~And \citep{vallerga93}.  An 18,000~K temperature would
suggest a \ion{Ca}{2} width of $\sim 2.7$ km~s$^{-1}$, which is much higher than the
1.5 km~s$^{-1}$ width reported by \citet{vallerga93}, albeit with
substantial uncertainty.

We conclude, therefore, that there is good reason to prefer a fit to the data
with an astrospheric component.  Figure~\ref{fit13}b shows such a fit, with the
Doppler parameters (i.e., temperatures) of the ISM components forced
to be consistent with best estimates inferred from the \ion{Mg}{2}, \ion{Fe}{2}, and
\ion{Ca}{2} lines, but with column densities allowed to vary.  The resulting
fit exhibits a slightly improved $\chi^2_{\nu}=1.19$, and with substantial
qualitative improvement in the $[-60,-50]$ km~s$^{-1}$ velocity range compared to
the ISM-only fit.  We conclude that an astrosphere
detection for $\lambda$ And is likely, consistent with the original assessment of \citet{wood96}; but this exercise clearly demonstrates how knowledge of the ISM
velocity structure can help illuminate the issues involved in making
such an assessment.

\subsection{Circumstellar Disks}
\label{sec:disks}

Narrow absorption features in stellar spectra usually are signatures of foreground interstellar absorption, but under certain circumstances, they could also result from circumstellar material, such as in a red-giant wind or the disk of a young star. Circumstellar disks, in particular, evolve through phases classified by their gas-to-dust ratio. Protoplanetary disks generally exist around pre-main sequence stars where accretion of material is ongoing, and the associated gas-rich disks are massive and optically thick. Transitional disks have optically thin inner regions and optically thick outer regions as revealed by mid- to far-IR excesses but little to no near-IR excess. Sub-millimeter CO emission indicates that these outer regions are gas rich \citep[e.g.,][]{qi04}.  Approximately 10$^7$ years into the star's lifetime, the primordial material clears, and the now main sequence star is surrounded by a gas-poor debris disk. Mechanisms that remove gas from the system include accretion, formation of a gas giant planet, depletion onto dust grains, and a wind from the central star.

UV and optical spectroscopy has been used to detect small amounts of gas in debris disks \citep[e.g.,][]{lagrange98, chen03}. This gas is not primordial, but rather results from collisions and evaporation of planetesimals \citep{roberge08}. Detecting gas in the debris disk of a star is challenging because by their nature very little gas exists in such disks. Sensitive observations of nearby, edge-on systems offer the best prospects for detecting absorption from disk gas in the UV and optical. The A-type star $\beta$ Pictoris, which satisfies these prerequisites, has become the iconic example of well-characterized gas absorption in a debris disk. UV and optical observations have indicated roughly solar abundances of gaseous elements with the exception of a large overabundance of carbon \citep{roberge06}. Similar characterizations of other circumstellar disk systems will enable better understanding of planet formation and composition.

\subsubsection{Absorption Measurements Towards Stars With Circumstellar Disks}

Two stars observed in the SNAP survey are known to possess circumstellar disks. Each of their spectra show multiple narrow absorption features. A true circumstellar absorption line should be centered at the radial velocity of the star. However, if nearby sight lines show a similar feature, or a kinematic model of the LISM predicts a cloud with the same projected velocity, then it is less likely that the absorption is circumstellar.  Although it could, of course, be an ISM/circumstellar blend. 


\subsubsubsection{49 Cet}

One particular sight line of interest is towards 49 Cet, an A1V star 59.4 pc away \citep{leeuwen07}. 49 Cet shows an infrared excess indicative of optically thin circumstellar dust grains \citep{sadakane86}. \citet{roberge13} presented far-infrared observations of dust emission and atomic gas emission lines obtained by the {\it Herschel Space Observatory}.  Molecular CO gas emission reveals that the disk is in the rare transitional phase in which the inner disk is cleared of molecular gas, while a significant quantity of gas is maintained in the outer disk \citep{zuckerman95,dent05,hughes08}.  \citet{hughes08} presented submillimeter CO spectra that show an extended molecular gas distribution approximately edge-on ($i = 90^{\circ} \pm 5^{\circ}$) with a heliocentric velocity of 12.2 $\pm$ 1.0 km s$^{-1}$.  Its near edge-on alignment makes it conducive to searches of gas absorption along the path length through the disk.  \citet{montgomery12} and \citet{welsh13} show temporal variability in \ion{Ca}{2} absorption for 49 Cet, and other A stars with circumstellar disks, possibly indicative of the evaporation of infalling exocomets.  


We detect two partially blended, narrow absorption features along the line of sight ($v_{\rm{MgII}}$ = 9.0 $\pm$ 1.3 km s$^{-1}$, 14.4 $\pm$ 1.1 km s$^{-1}$ ; $v$$_{\rm{FeII}}$ = 11.0 $\pm$ 1.6 km s$^{-1}$, 13.65 $\pm$ 0.15 km s$^{-1}$). Given the presence of an edge-on, optically thin, gas-rich disk around 49 Cet, it is possible that the disk produces one of the components. The \citet{redfield07lism4} kinematic model predicts that this sight line traverses only one cloud, the LIC with $v_{\rm{LIC}} = 11.00 \pm 1.29$ km s$^{-1}$.  While component 1 ($\sim$10 km~s$^{-1}$) is consistent with the LIC velocity at the 0.6$\sigma$ level, component 2 ($\sim$14 km~s$^{-1}$) is discrepant by 1.4$\sigma$.  Unfortunately, the stellar rest frame is also at a very similar velocity (12.2 km~s$^{-1}$; \citealt{hughes08}).  Component 2 is consistent with the stellar rest frame at the 0.9$\sigma$ level, while component 1 is discrepant at the 1.7$\sigma$ level.  While it is not possible to make a definitive conclusion, we find that component 1 is most likely associated with the LIC and component 2 is associated with the disk.  A reliable test to differentiate LISM from disk absorption is to observe other nearby sight lines to confirm that the LISM absorption remains, but the disk absorption is absent.  Unfortunately, the nearest sight lines to 49 Cet are $>$10$^{\circ}$ away ($\beta$ Cet, 12.4$^{\circ}$, 29.5 pc; and $\sigma$ Cet, 13.8$^{\circ}$, 26.7 pc).  This is a complicated region of the sky, with many different LISM clouds in the vicinity: $\beta$ Cet shows absorption from the Mic and Cet Clouds and $\sigma$ Cet shows absorption from the LIC, Blue, and G Clouds.  Observations of significantly closer sight lines at similar distances to 49 Cet are needed to clearly distinguish the LISM absorption from any disk absorption.

If we assume that component 2 of 49 Cet is indeed caused by disk absorption, then we can compare the observed column densities with another edge-on system, $\beta$ Pic.  $\beta$ Pic has a comprehensive elemental inventory of atomic absorption line detections compiled by \citet{roberge06}, including the three ions observed in the SNAP sample, \ion{Mg}{2}, \ion{Fe}{2}, and \ion{Mn}{2}.  \citet{lagrange98} measured column densities for all three ions, $\log N($\ion{Mg}{2}$) \geq$ 13.3, $\log N($\ion{Fe}{2}$) = 14.5$, $\log N($\ion{Mn}{2}$) = 12.5$.  The column densities derived here for 49 Cet are all an order of magnitude smaller than for $\beta$ Pic, perhaps due to a slight difference in inclination between the two.  The smaller column density observed toward 49 Cet might result from $\beta$ Pic being viewed edge-on, while the line of sight to 49 Cet is more grazing and traverses the more tenuous atmosphere of the disk.  Indeed, substantial atomic absorption line variability is detected for $\beta$ Pic \citep{kondo85,ferlet87,petterson99}.  \citet{montgomery12} monitored 49 Cet in \ion{Ca}{2} and found evidence for variability, but at levels significantly below that seen in $\beta$ Pic.  They also measured two stable components at heliocentric radial velocities of 10.3 and 15.0 km~s$^{-1}$, consistent with our UV observations.  

\subsubsubsection{HD141569}

The sight line towards HD141569 is the second longest sight line in the sample (116 pc). As expected, the absorption in this direction is deep and complicated. Six components were observed in \ion{Mg}{2}, four of which are saturated, as well as five components in \ion{Fe}{2} and four in \ion{Mn}{2}. HD141569 is a B9.5V Herbig Ae/Be star \citep{jaschek92}. Spectral energy distribution (SED) observations indicate a large ($\sim$400 AU) circumstellar disk inclined by 51 $\pm$ 3$^\circ$  \citep{weinberger99}.  \cite{dent05} measured a radial velocity of $-$7.6 $\pm$ 0.3 km s$^{-1}$ using the disk's double-peaked CO $J = 2 - 1$ spectral profile.

The \citet{redfield07lism4} kinematic model predicts that only one LISM cloud should be traversed, the G Cloud.  The radial velocity of component 2 matches that of the G cloud to within 0.7$\sigma$.  Component 3 also is consistent with the Leo Cloud to within 0.4$\sigma$, for which the boundary is $<$20$^{\circ}$ away.  The remaining four absorption components are unassociated with any of the known LISM clouds within 15 pc.  They might be caused by more distant interstellar structures.  However, component 5, with a radial velocity of $-6.0 \pm 1.1$ km~s$^{-1}$ is consistent with the stellar rest frame at the 1.4$\sigma$ level, and could be due to the circumstellar disk.  The orientation of the system is not edge-on, which raises concerns whether any absorption would be detectable, although an inner gas disk with an extended atmosphere may be a possibility.  Other nearby sight lines of close angular distance should be observed before determining whether the HD141569 disk indeed is detected in absorption.


\subsubsection{Absorption Measurements Along Sight Lines Near a Star With a Circumstellar Disk}

Two sight lines in the SNAP survey probe the region close to HD32297, an A0V star at 112 pc with an edge-on debris disk \citep{schneider05}. \cite{redfield07} observed \ion{Na}{1} absorption lines in the optical toward HD32297 and several nearby stars, all of which show a LIC absorption component. Only for HD32297, however, was a second absorption component detected, consistent with the $\sim$20 km s$^{-1}$ stellar radial velocity. Five observations of HD32297 over 5 months show the same two components, confirming that this unique component is \ion{Na}{1} gas absorption in the stable edge-on debris disk.  \citet{debes09} provided an explanation for the asymmetries in the scattered light images of HD32297 by appealing to a model accounting for the interaction of the disk with its surrounding interstellar medium.

Both HD28911 (9.0$^\circ$ separation from HD32297) and 85 Tau (11.2$^\circ$ separation) show similar absorption components (see Table~\ref{32297table}), confirming the \cite{redfield07lism4} kinematic model prediction that three discrete clouds (LIC, Aur, Hyades) are present along their lines of sight (see Section~\ref{secsss}).  The kinematic model predicts LIC absorption toward HD32297 at a velocity of $23.6$ km~s$^{-1}$.  \citet{fusco13} analyzed UV absorption lines detected toward HD32297 for several ions (\ion{Mg}{1}, \ion{Mg}{2}, \ion{Mn}{2}, \ion{Zn}{2}, \ion{Fe}{1}, and \ion{Fe}{2}).  All of these lines show absorption at $\sim$23.2 km~s$^{-1}$, consistent with the LIC prediction.  All but \ion{Mn}{2} also show absorption at $\sim$19.1 km~s$^{-1}$, consistent with the $\sim$20 km s$^{-1}$ disk absorption detected in \ion{Na}{1} by \citet{redfield07} and with the stellar radial velocity.  In the directions of HD28911 and 85 Tau, the Hyades Cloud is definitely detected near the predicted velocity of 13.2 km~s$^{-1}$ and the Aur Cloud may be detected, although the observed radial velocities are about 2.2 km~s$^{-1}$ less than the predicted velocities.  For HD32297, the Aur Cloud with a predicted velocity of 23.9 km~s$^{-1}$ may contribute to the 23.2 km~s$^{-1}$ absorption feature ascribed to the LIC, but it is unlikely that the 20.5 km~s$^{-1}$ absorption is due to the Aur Cloud.  The new SNAP observations support the claim that the $\sim$19.1 km~s$^{-1}$ absorption detected in HD32297 is due to the edge-on disk, and that the boundaries of the Hyades and Aur Clouds do not extend to the sight line toward HD32297.

\begin{deluxetable} {lccclcccc}
\tablewidth{0pt}
\tabletypesize{\scriptsize}
\tablecaption{Comparison of LISM Absorption in Direction of HD32297 \label{32297table}}
\tablehead{Star & \multicolumn{2}{c}{Galactic Coordinates} & Distance & Observed Absorption & Stellar Radial & \multicolumn{3}{c}{Predicted Cloud Velocities}\\
 & $l$ & $b$ & & Velocities & Velocity & LIC & Hyades & Aur \\
& (deg) & (deg) & (pc) & (km s$^{-1}$) & (km s$^{-1}$) & (km s$^{-1}$) & (km s$^{-1}$) & (km s$^{-1}$)}
\startdata
HD32297 & 192.8 & --20.2 & 112  & 24.4, 20.5 (NaI)\tablenotemark{a} & ~20  & 23.6 & 12.3 & 23.9\\
        &       &        &      & 23.2, 19.1 (UV lines)\tablenotemark{b} & &    &      & \\
HD28911 & 183.4 & --22.6 & 44.7 & 23.8, 14.3, 20.3 & 35 & 23.5 & 13.2 & 22.4\\
85 Tau  & 180.9 & --21.4 & 45.2 & 23.3, 13.8, 18.6 & 36 & 23.5 & 13.3 & 21.9\\
\enddata
\tablenotetext{a}{\citet{redfield07hd32297}}
\tablenotetext{b}{\citet{fusco13}}
\end{deluxetable} 



\section{Conclusions}
\label{conclusion}

High-resolution NUV SNAP observations of 34 stars within $\sim$100 pc broadly distributed across the sky reveal widespread \ion{Mg}{2}, \ion{Fe}{2}, and \ion{Mn}{2} absorption in the LISM. Among these sight lines, we detected 76 \ion{Mg}{2} components, 71 \ion{Fe}{2} components, and 11 \ion{Mn}{2} components. Each \ion{Fe}{2} and \ion{Mn}{2} component matches an \ion{Mg}{2} component to within 3 km s$^{-1}$ in radial velocity, evidence that they arise from the same LISM clouds. The distribution of radial velocities is consistent with the bulk flow of the cluster of local interstellar clouds, and the Doppler parameters reflect the greater contribution of thermal broadening for the lighter \ion{Mg}{2} ion. The average number of components per sight line remains flat beyond 10 pc, and only begins rising beyond $\sim$60 pc, evidence that LISM clouds are concentrated close to the Sun, and that a considerable accumulation of material traces the edge of the Local Bubble.

Every prediction made by the \cite{redfield07lism4} kinematic model of the LISM is confirmed by an observed component in the new lines of sight. The success of the model points to the value of these observations for understanding the velocity structure of the LISM. Many velocity components not predicted by the model for the observed lines of sight agree with the projected velocities of nearby clouds within small angular separations from these lines of sight. In these cases, the cloud boundaries will need to be redrawn. For longer lines of sight, we detected many absorption components not consistent with previously identified clouds.  These can be compared with unidentified components along nearby sight lines to construct velocity vectors for new clouds.

Three close pairs of sight lines in this sample, separated by $<$3$^{\circ}$, were scrutinized for evidence of small scale structure.  While slight variations in the absorption profiles can be seen, we find little evidence for significant small scale structure.  For the dominant ions, the radial velocities, Doppler widths, and column densities are consistent for scales on the order of 10,000 AU or 0.05 pc.

All of the new NUV SNAP spectra are along sight lines with existing FUV spectra, typically taken at medium resolution because of the low intrinsic stellar flux.  The new high-resolution NUV spectra provide critical information regarding the velocity structure of the LISM absorbers, which can then be applied to the blended medium-resolution spectra of the lighter ions to deduce fundamental physical properties.  For example, the widths of LISM absorption lines contain information on the temperature and turbulent velocity of the gas.  We use the line widths measured in this paper, together with archived FUV absorption lines to make the first measurement of temperature and turbulent velocity along the sight line to $\epsilon$ Ind.  The absorption is associated with the LIC, and the temperature and turbulent velocity are consistent with other measurements of the LIC.  Accumulating a large sample of these combined NUV $+$ FUV measurements will be critical in evaluating the homogeneity of LISM clouds.

Clouds detected towards $\epsilon$ Ind and $\lambda$ And are of particular interest because these stars show evidence of astrospheres. Understanding the LISM cloud velocity structure in the foreground of these stars influences the fitting of the often blended and saturated Ly$\alpha$ lines used to detect the subtle astrospheric absorption. The $\epsilon$ Ind sight line shows evidence for only one interstellar cloud, which was assumed in the original Ly$\alpha$ fitting. On the other hand, we detected three components toward $\lambda$ And, which was previously modeled assuming only a single high temperature cloud. We analyzed the Ly$\alpha$ line with three ISM clouds included in the fit. An astrosphere detection remains viable in our analysis, which highlights the importance of high-resolution LISM spectra to constrain the velocity structure of the interstellar absorption.

Two stars in the sample have known circumstellar disks. 49 Cet, which has an edge-on debris disk, shows \ion{Mg}{2}, \ion{Fe}{2}, and \ion{Mn}{2} absorption at the stellar velocity; and likewise for the \ion{Mg}{2} and \ion{Fe}{2} components towards HD141569. In both sight lines, the proposed disk components do not agree with any model LISM cloud predictions. Eliminating the possibility that this absorption is from the ISM would require further examination of nearby sight lines.  Such an analysis of two nearby sight lines (HD28911 and 85 Tau) in the SNAP sample to the sight line toward HD32297, an edge-on debris disk located 112 pc away, supports the detection of disk absorption in the HD32297 spectrum.  

The results presented here are only the beginning in a series of investigations that will characterize the LISM and its constituent clouds. When combined with archived FUV spectra, often obtained at medium resolution, it will be possible to measure the temperature and turbulence of LISM clouds as was shown with the LIC towards $\epsilon$ Ind. Furthermore, observations of different ionization stages of Mg, Fe, and Mn along the same sight lines can help describe the interstellar radiation field. Similarly, a comparison of column densities of various ions, across many sight lines, provides a valuable inventory of the abundances and depletions of LISM clouds. With more sight lines, tighter constraints can be placed on the three dimensional morphology of the LISM, including the small scale structure of the clouds.  Due to the proximity of LISM clouds, a large number of sight lines must be observed in order to adequately sample the cloud structures.  As this sample grows, it will be possible to integrate the various fundamental measurements into a self-consistent model of the morphology and physical characteristics of the structures inhabiting our local interstellar environment.

\acknowledgments
We would like to thank the anonymous referee for their insightful comments.  We thank Meredith Hughes for several helpful discussions and useful comments.  We acknowledge support through NASA HST Grant GO-11568 from the Space Telescope Science Institute, which is operated by the Association of Universities for Research in Astronomy, Inc., for NASA, under contract NAS 5-26555; and a student research fellowship from the Connecticut Space Grant Consortium.

{\it Facilities:} \facility{HST (STIS)} \facility{}

\clearpage

\end{document}